\documentclass[aps,pre,twocolumn,eqsecnum,superscriptaddress]{revtex4-1}

\usepackage{color}
\usepackage{hyperref}
\usepackage{amsmath}
\usepackage{graphicx}
\usepackage{color}
\usepackage{wrapfig}
\usepackage[export]{adjustbox}
\begin{document}


\title{The Spanning Tree Model and the Assembly Kinetics of RNA Viruses.}


\author{Inbal Mizrahi}
\affiliation{Department of Physics and Astronomy, University of California, Los Angeles, CA 90095}
\author{Robijn Bruinsma}
\affiliation{Department of Physics and Astronomy, University of California, Los Angeles, CA 90095}
\affiliation{Department of Chemistry and Biochemistry, University of California, Los Angeles, CA 90095}
\author{Joseph Rudnick}
\affiliation{Department of Physics and Astronomy, University of California, Los Angeles, CA 90095}


\date{\today}

\begin{abstract}
Single-stranded (ss) RNA viruses self-assemble spontaneously in solutions that contain the viral RNA genome molecules and viral capsid proteins. The self-assembly of \textit{empty} capsids can be understood on the basis of free energy minimization. However, during the self-assembly of complete viral particles in the cytoplasm of an infected cell, the viral genome molecules must be selected from a large pool of very similar host messenger RNA molecules and it is not known whether this also can be understood by free energy minimization. We address this question using a simple mathematical model recently proposed for the assembly of small ssRNA viruses \footnote{submitted to PLOS Biocomputation.}. We present a statistical physics analysis of the properties of the model finding an effect kinetic RNA selection mechanism with selection taking place during the formation of the nucleation complex. Surprisingly, kinetic selectivity is greatly enhanced by a modest level of supersaturation and by reduced protein to RNA concentration ratios. The mechanism is related to the Hopfield kinetic proofreading scenario.
\end{abstract}


\maketitle

\section{Introduction} \label{sec:intro}

Many single-stranded (ss) RNA viruses, such as the polio and common cold viruses, are able to self-assemble spontaneously into infectious viral particles (``virions") in solutions that contain appropriate concentrations of viral capsid proteins and RNA molecules \cite{Fraenkel-Conrat1955,Butler1978}. For these viruses, assembly is believed to be a purely passive process driven by free energy minimization. Early work by Aaron Klug \cite{Klug1999} indicated that RNA genome molecules act as \textit{templates} that direct the viral assembly process. He proposed a physical model for viral assembly in which the repulsive electrostatic interactions between positively charged groups of the capsid proteins are just strong enough to overcome competing attractive hydrophobic interactions between the proteins, thus preventing the self-assembly of empty capsids under physiological conditions. When viral RNA molecules are then added to the solution, the negative charges of the RNA nucleotides neutralize some of the positive charges of the capsid proteins thereby tilting the free energy balance towards assembly \cite{Schoot2005, Forrey2009, Garmann2014a, Perlmutter2015b},\footnote{For a quantitative treatment, see ref. \cite{Kegel2006}}. 

 Viral gRNA molecules must compete for packaging with a large pool of -- quite similar -- host messenger RNA (mRNA) molecules for packaging by the viral capsid proteins \cite{Dimmock2001}. For the case of influenza the number of gRNA molecules inside an infected cell is less than $10^4$ \cite{frensing} while the total number of host mRNA molecules is in the range of $3.6\times 10^5$. For the HIV-1 virus, the number of gRNA molecules may be as low as $10^2$. Like other ssRNA molecules, genomic RNA molecules (gRNA) have a tree-like ``secondary structure" produced by Watson-Crick base-pairing between complementary RNA nucleotides of the primary sequence of RNA nucleotides \cite{mathews1999}. The redundancy of the genetic code allows for the possibility of ``silent'' (or synonymous) mutations that can alter the secondary structure of the molecule without altering the structure of the proteins encoded by the nucleotide sequence \cite{tubiana}. Viral gRNA molecules appear to have undergone different forms of evolutionary adaptation increasing the packaging probability. On the one hand, they have short specific sections, known as \textit{Packaging Signals} (PS) that have a specific affinity for the capsid proteins of the virus \cite{Frolova1997, Basnak2010, Bunka2011, Stockley2013a, Dykeman2013a, Dykeman2013b, Patel2015}. On the other hand, the global topology of gRNA molecules differs from that of generic mRNA molecules: they are longer and significantly more branched and compact, which reduces the radius of gyration of the RNA molecules in solution and hence the free energy cost of compacting the RNA molecules prior to encapsidation \cite{yoffe2008}. 

The physical aspects of ssRNA packaging have been extensively studied experimentally, theoretically, and by numerical studies of model systems \cite{Zhang2004, Schoot2005, Kegel2006, Belyi2006, Hu2008a, Devkota2009, Hagan2009, Forrey2009, Jiang2009, Siber2010, Ting2011, Ni2012, Siber2012, Ford2013, Zhang2013c, Erdemci-Tandogan2014, Garmann2014b, Kim2015, bond2020}, \footnote{For reviews, see \cite{roos2010, zandi2020, bruinsma2021}}. The theoretical studies focused largely on the minimization of the free energy of assembled virions, which produced global measures for the packaging fitness of ssRNA molecules in terms of their length and compactness. On the other hand, experimental studies of the self-assembly of \textit{empty} capsids \cite{Prevelige1993, Casini2004, medrano} were interpreted in terms of a kinetic nucleation-and-growth scenario, where the energetically uphill formation of a ``nucleation complex", composed of a small number of capsid proteins, is followed by an energetically downhill ``elongation process" that ends with the closure of the capsid. This nucleation complex may be compared to the critical nucleus of the classical theory of nucleation and growth theory as applied to empty capsid assembly \cite{Zandi2006, bruinsma2021}. The TMV assembly scenario proposed by Klug is in fact an example of a nucleation and growth scenario with a nucleation complex composed of a single PS associating with single disk of proteins with the subsequent elongation proceeding by the addition of additional disks. The contribution of a single PS to the assembly free energy could be extremely small, suggesting that free energy minimization may not be suitable to understand RNA selection. Recent observations on the assembly kinetics of individual MS2 viruses (a small ssRNA bacteriophage virus) reported a wide distribution of time scales \cite{garmann2019}, which is what is expected from a nucleation-and-growth scenario. Next, for the case of the assembly of the HIV-1 retrovirus (see ref.\cite{comas2017} and references therein), RNA selectivity has been shown to depend on the cooperative action of a cluster of PS located at the 5' end of the gRNA molecule, known as the $\psi$ sequence. This sequence is about a hundred nucleotides long, small compared to the total length of the HIV-1 genome of about $10^4$ nucleotides. HIV-1 gRNA selection appears to take place during the nucleation stage of the assembly process when this $\psi$ sequence interacts with a small group of capsid proteins. Changing the RNA sequence of the non-$\psi$  of the genome molecules does not affect the selectivity. The PS of HIV-1 was shown to provide no significant thermodynamic advantage to the gRNA molecules over non-viral RNA molecules of the same length\cite{jouvenet}. As for the MS2 case, the HIV-1 assembly process is characterized by a broad distribution of time scales.

\begin{figure}[htbp]
\begin{center}
\includegraphics[width=2.5in]{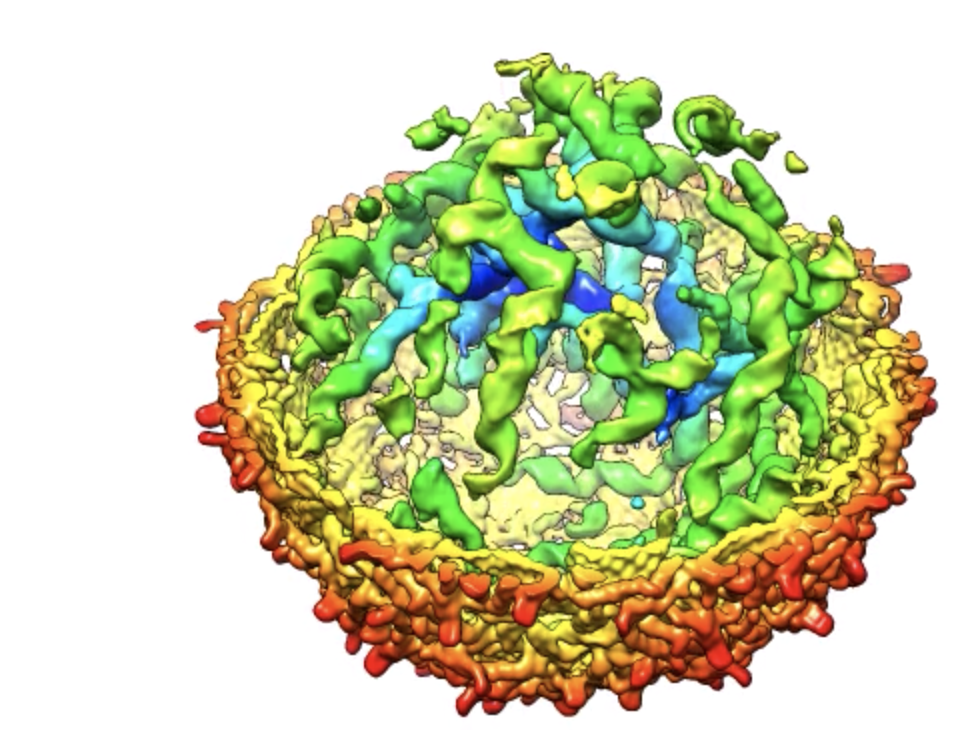}
\caption{CryoEM asymmetric reconstruction of the MS2 virion. The viral genome molecules (in green and blue) associate reproducibly mostly with one half of the capsid shell. From ref.\cite{dai2017}}.
\label{fig:MS2}
\end{center}
\end{figure}

Important information about the kinetics of viral co-assembly can be gleaned as well from purely structural studies. Until recently, reconstruction of packaged genome molecules involved ``icosahedral averaging", which resulted in RNA structures with imposed icosahedral symmetry \cite{Baker}. Such studies showed that the interior surface of the icosahedral capsids of the nodaviruses \cite{Tihova2004, Johnson2004}) is decorated by paired RNA strands lining the edges of the ``capsomers" (i.e., pentameric or hexameric groupings of capsid proteins). Recent progress in cryo-electron tomography has made it possible to reconstruct the way individual ssRNA genome molecules are packaged inside spherical capsids without having to resort to icosahedral averaging (``asymmetric reconstruction" \cite{koning, beren}). An important example is again the MS2 virus. It was found that sections of the RNA genome rich in PS reproducibly associate with roughly \textit{half} of the interior surface of the capsid \cite{Dykeman2011, dai2017}, as shown in Fig.\ref{fig:MS2}
The remaining capsid proteins do not associate in a reproducible manner with the gRNA. This can be interpreted as evidence for a critical nucleus in the form of a well-defined nucleo-protein complex held together by a particular section of the viral RNA molecule that is rich in packaging sequences. The subsequent downhill elongation process is driven by generic electrostatic interactions.

These observations indicate that, at least for the TMV, MS2 and HIV-1 viruses, RNA selection by PS is best understood as a \textit{kinetic} effect where the PS reduce the height of the assembly energy activation barrier. This selectively increases the assembly rate of viral particles with respect to the packaging of host mRNA molecules. Since the production rate of virions depends \textit{exponentially} on the height of activation energy barrier, a limited number of PS could have a disproportionally large effect. This mechanism might be called ``selective nucleation"\footnote{Selective nucleation was proposed by I. Rouzina in the context of the assembly of retroviruses}. The action of PS would be similar to that of enzymes or catalysts that increase the rate of a chemical reaction by reducing the height of an energy activation barrier. 

It should be emphasized that these observations are not inconsistent with RNA compaction also playing a central role in viral assembly. Suppose the activation energy barrier of a particular RNA molecule happens to be unusually low, so it is selected during the nucleation stage. This molecule will not actually get packaged if its size is be too large to fit inside the capsid. One might speculate that the PS operate on the level of the nucleation step during the early stages of assembly kinetics while RNA compactness operates on the level of the elongation step. In actuality we will see that compactness also can play an important role during the nucleation stage. The aim of this article is to explore the general physical aspects of kinetic RNA selection by PS control of the assembly activation energy. The questions we propose to answer are: what is the maximum kinetic specificity and how long can this kinetic specificity persist assuming selectivity in the face of full thermal equilibriation? Next, what are the general features of the selection process in terms of the assembly speed, the distribution of time scales and how do these quantities depend on the RNA geometry and topology? We will use a recent mathematical model \footnote{A short account of the model has been submitted to Plos Computational Biology}, the \textit{``Spanning Tree Model"}, to address these questions. This model includes tens of thousands of RNA secondary structure configurations on the a compacted RNA core. All these configuration have the same final assembly energy but they can have have different assembly energy barriers and different numbers of assembly pathways. In this model, there is practically no selectivity under conditions of thermodynamic equilibrium so it allows to focus on selectivity generated by the nucleation barrier. The model, which is sufficiently simple so its kinetics can be determined by numerical integration of a set of coupled Master Equations, is itself a generalization of an earlier model for the assembly of \textit{empty} dodecahedral capsids by Zlotnick \cite{Zlotnick1994, Endres2002, Zlotnick2007}. The kinetics of the Zlotnick model obeys a nucleation-and-growth assembly scenario \cite{Morozov2009} and it also has been used to carry out simulations of the packaging of linear genome molecules \cite{Perlmutter2014, Perlmutter2015b}. 

The spanning tree model is introduced in Section II, followed by a topological and geometrical classification of the model genome molecules. Next is a discussion of minimum energy assembly pathways and of the structural properties of the partial assemblies. In section III, a non-linear Master Equation for the assembly kinetics is introduced. Numerical integration of this Master Equation is used to determine the characteristic time scales of the assembly kinetics and to study packaging competition between different classes of genome molecules as well for different levels of supersaturation and RNA-to-protein mixing ratios. In Section IV we examine two-stage packaging scenario in order to compare the packaging selectivity of protein-by-protein assembly kinetics with collective assembly kinetics, such as the \textit{en-masse} scenario \cite{Perlmutter2014,Perlmutter2015}. In the concluding Section V we summarize our results, discuss experimental predictions and limitations of the model that could be improved upon.

\section{The Spanning Tree Model.}

\subsection{Empty Capsid Assembly}
\begin{figure}[htbp]
\begin{center}
\includegraphics[width=2.5in]{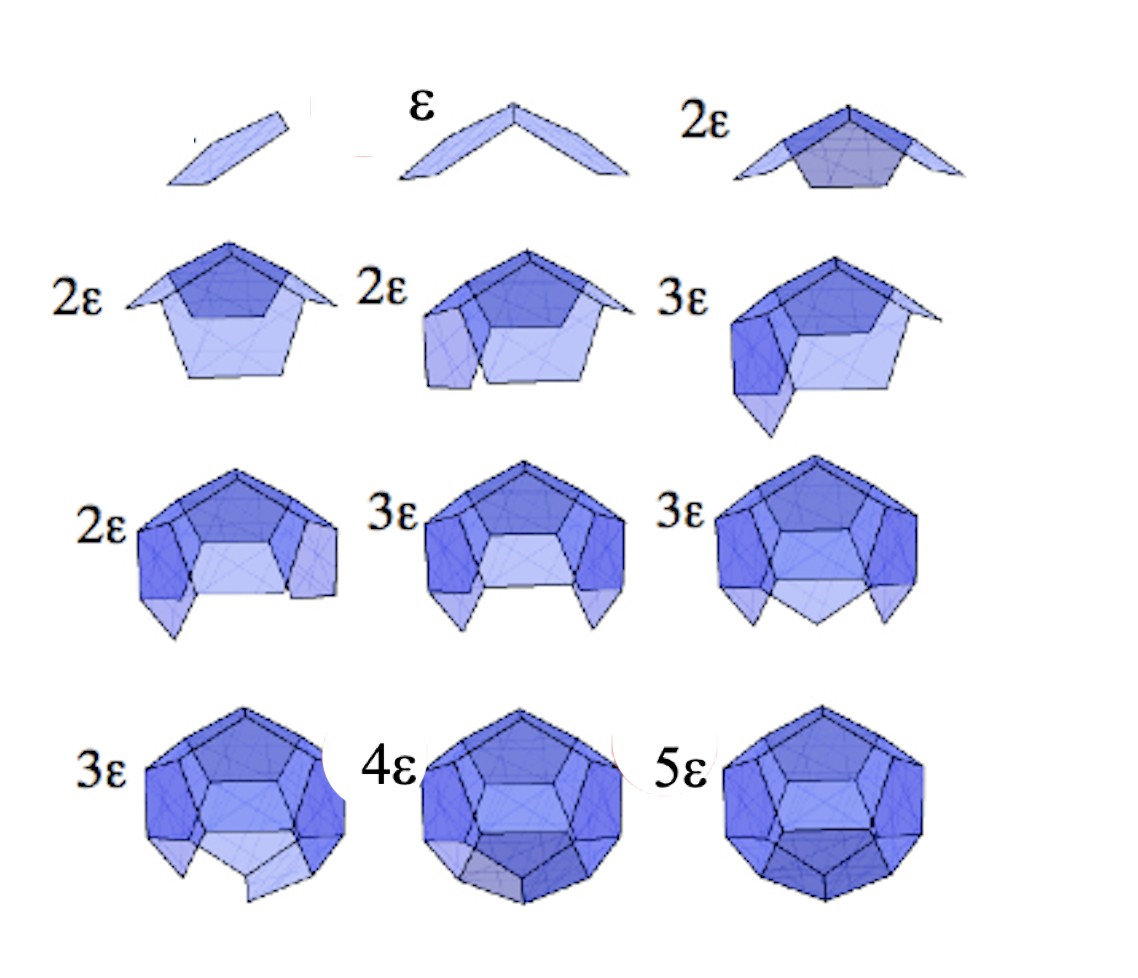}
\caption{Empty capsid assembly pathway. The figure shows one of the minimum-energy pathways for the assembly of a dodecahedral shell composed of twelve pentamers with adhesive edges. The edge-to-edge binding energy is $\epsilon$. The change in the total energy of the cluster for each added pentamer is indicated. Note that the assembly intermediates all are compact structures.}
\label{fig:Zlotnick model}
\end{center}
\end{figure}

The Zlotnick model treats the capsid as a dodecahedral shell composed of twelve pentamers.  The sixty proteins of the shell correspond to the capsid of a minimal ``T=1" virus. Assembly is driven by attractive edge-edge interactions between the pentamers. A \textit{minimum-energy assembly pathway} can be defined as a pentamer-by-pentamer addition sequence where each added pentamer is placed in a location that minimizes the free energy of the partial shell. An example of one of the very many ($\simeq10^5$) degenerate minimum-energy assembly pathways is shown in Fig.\ref{fig:Zlotnick model}.   
The assembly energy $\Delta E(n)$ of a partial shell composed of $n$ pentamers is defined to be $\Delta E(n)/E_0= - n_1 - n \mu_0$ with $n_1$ the number of shared pentamer edges of the partial shell and with $ E_0$ the edge-to-edge binding energy. This binding energy can be estimated by comparison with thermodynamic assembly studies of empty capsids, which gave values of about $4.3 k_bT$ \cite{Zlotnick1994}. In the following, all quantities with dimensions of energy will be expressed in units of the energy scale $E_0$. Next, $\mu_0$ is the pentamer chemical potential at a reference concentration. The assembly energy of a complete capsid equals $-30-12 \mu_0$ for all minimum energy assembly pathways. Assembly equilibrium is the state where the chemical potential of a pentamer in solution is the same as the energy of a pentamer that is part of a capsid. This is the case if $\Delta E(12)=0$ so for a reference chemical potential $\mu^*=-5/2$.  

\begin{figure}[htbp]
\begin{center}
\includegraphics[width=3.5in]{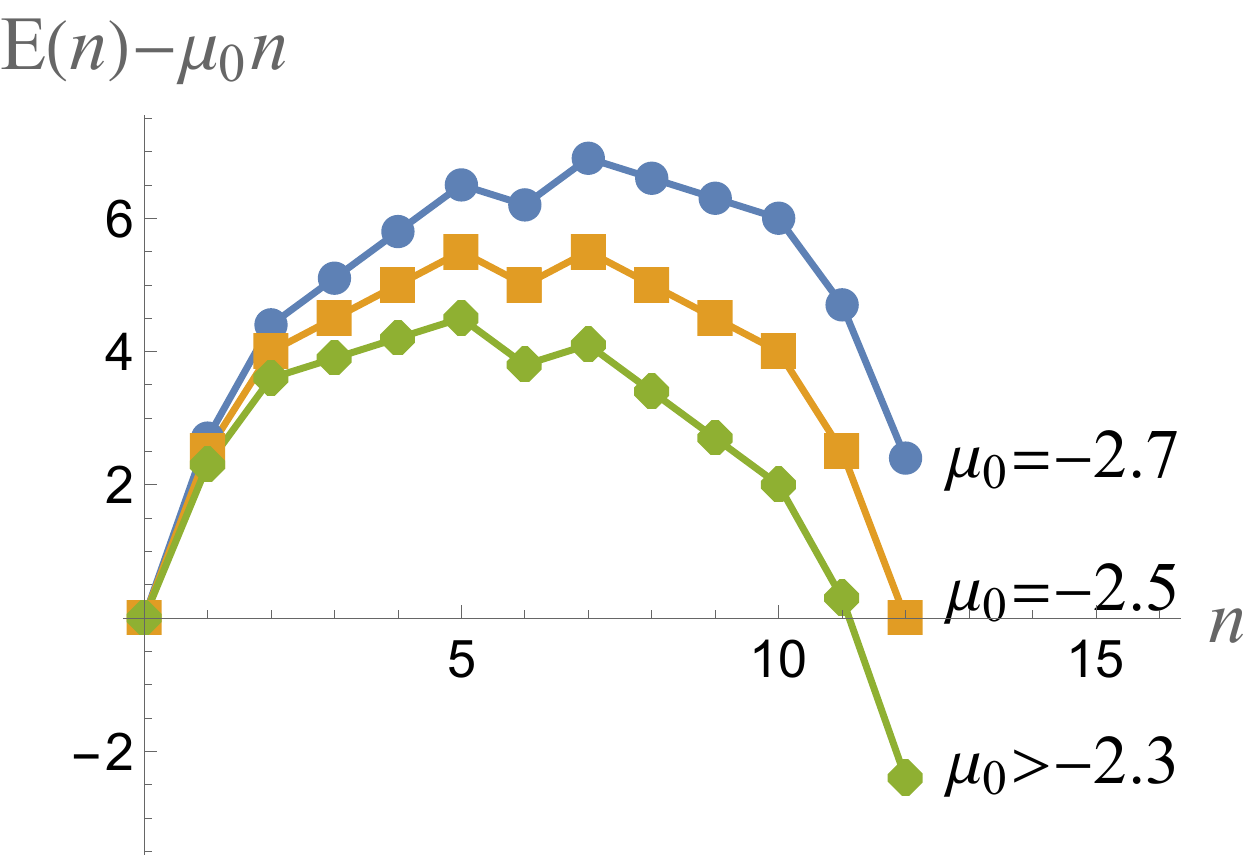}
\includegraphics[width=3.5in]{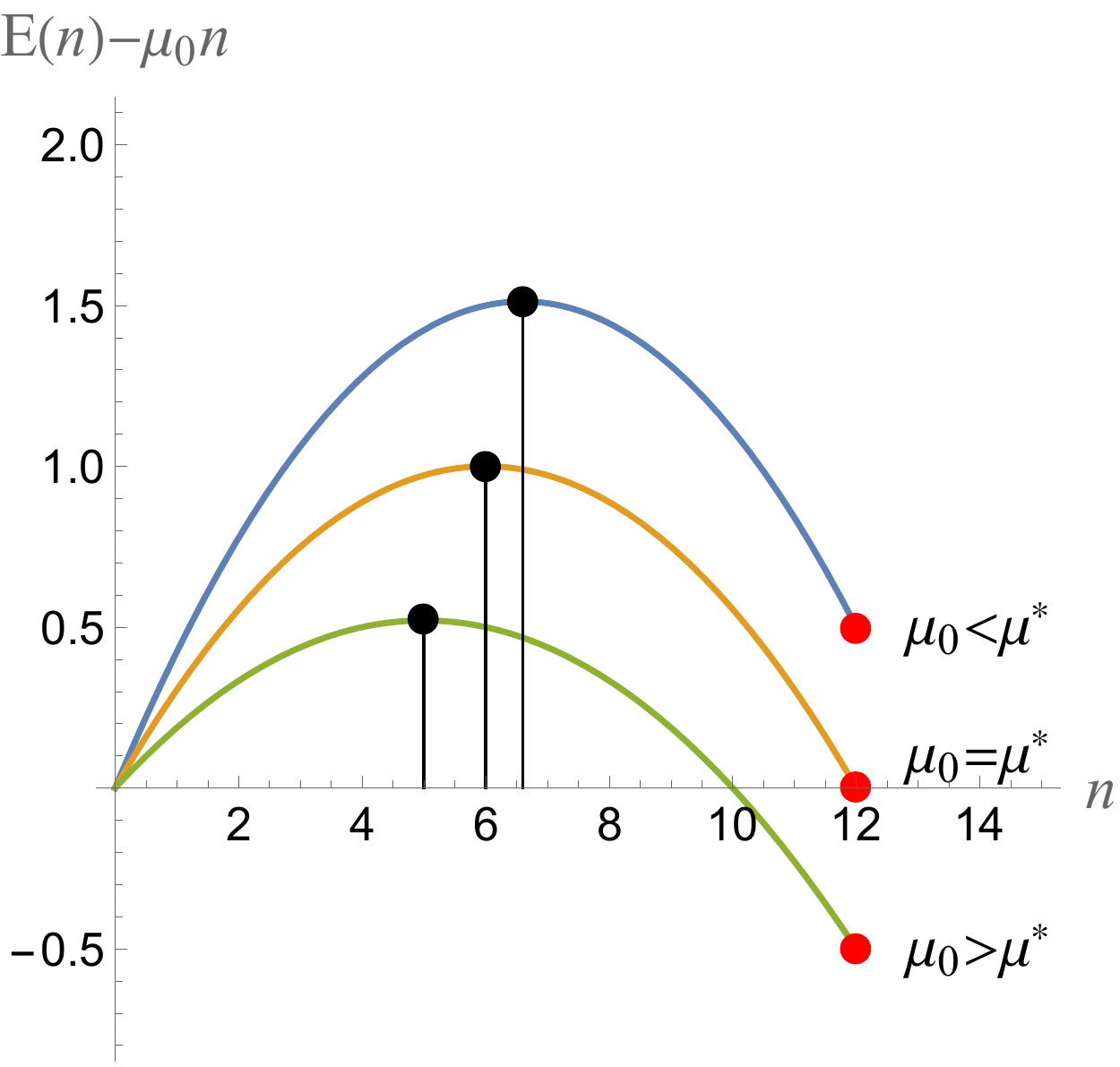}
\caption{Top: Energy profiles of a minimum energy assembly pathway of the Zlotnick model. Blue dots: The chemical potential $\mu_0$ is slightly below $\mu^*$, the value of the chemical potential for assembly equilibrium. Orange squares: $\mu_0$ is equal to $\mu^*$. Green diamonds: $\mu_0$ is slightly above $\mu^*$. Bottom: Assembly energy profiles according to the continuum theory of nucleation and growth  \cite{Zandi2006}. Solid red dots: energy minima. Solid black dots: energy maxima.  }
\label{fig.2}
\end{center}
\end{figure}

Figure \ref{fig.2}(top) shows the minimum energies of the n-pentamer partial assemblies of Fig.\ref{fig:Zlotnick model} for three different values of the reference chemical potential near $\mu^*$.
For $\mu_0<\mu^*$, the absolute energy minimum is at $n=0$ while for $\mu_0>\mu^*$ the absolute minimum is at $n=12$, the assembled capsid. The assembly activation energy barrier of a profile is the height of the maximum. The location $n^*$ of the maximum under equilibrium conditions corresponds to a half-filled shell, shifting to lower values as $\mu_0$ increases. For comparison, Figure \ref{fig.2} (bottom) shows the assembly energy of a spherical cap growing into a spherical shell \cite{Zandi2006}. The initial rise of $\Delta E(n)$ with $n$ is due to the fact that the line energy of the perimeter of the cap increases with $n$ for $n$ less than six while the subsequent drop of $\Delta E(n)$ is due to the fact that the line energy decreases as a function of $n$ for $n$ larger than six, when the perimeter starts to shrink. This plot resembles the ``classical" nucleation-and-growth theory form, shown in the  bottom figure.

\subsection{Spanning Trees and their Classification.}
\begin{figure}[htbp]
\begin{center}
\includegraphics[width=3in]{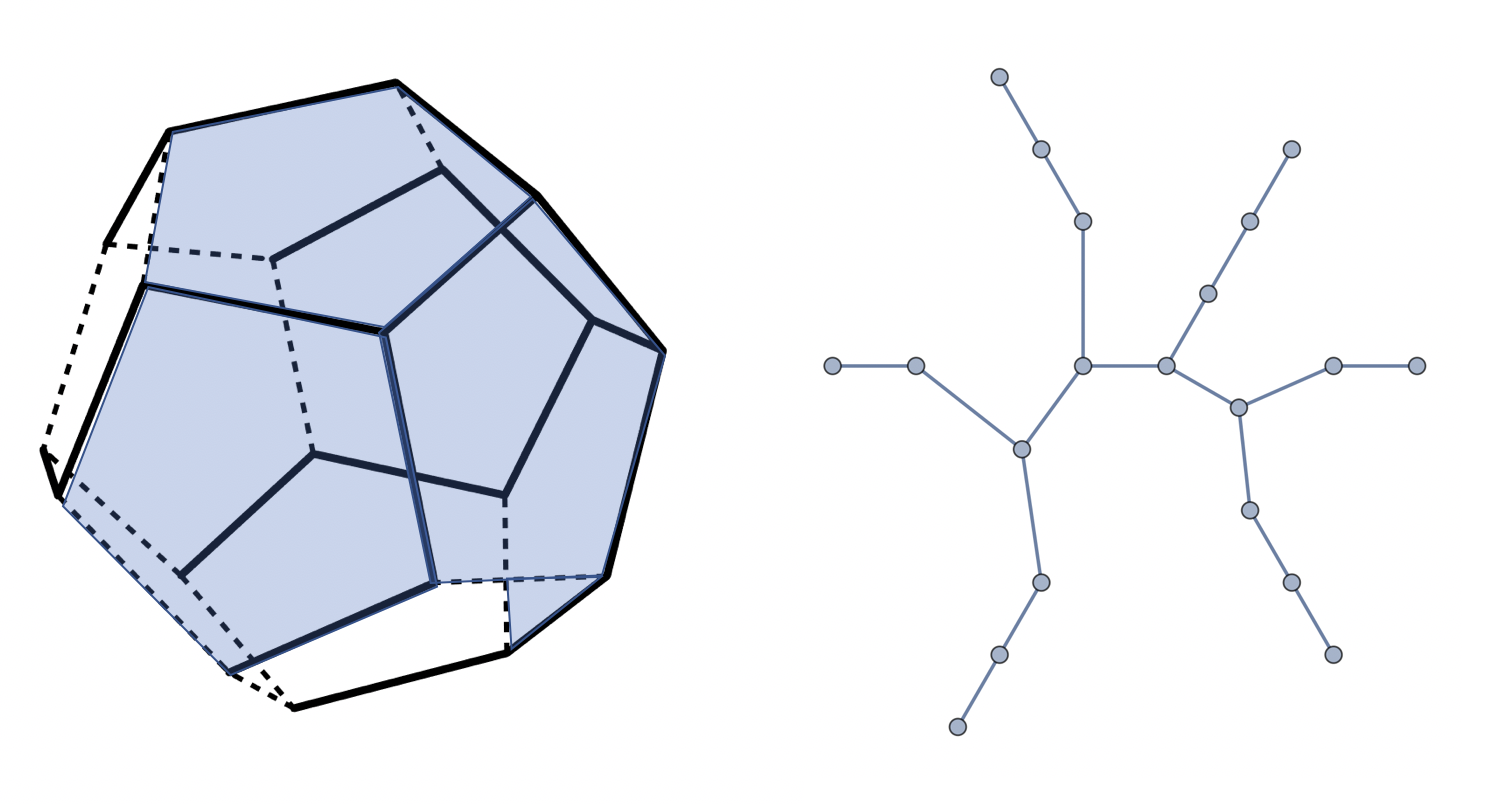}
\caption{Left: Spanning tree connecting the vertices of a dodecahedron (solid lines). The dashed lines indicate edges of the dodecahedron that are not a part of the spanning tree. Six pentamers, shown in blue, can be placed on the dodecahedron with each pentamer wrapped by four links of the tree. Right: Planar graph of the same spanning tree.}
\label{fig:dodtrees}
\end{center}
\end{figure}

The second part of the definition of the model concerns the representation of RNA molecules. The RNA molecules are assumed to have the same length. Prior to assembly, the molecules are assumed to be compacted into dodecahedra whose shape matches the interior of the dodecahedral capsid of the Zlotnick model. The molecules differ only in terms of a PS section that is in contact with the capsid. This section is assumed to have a secondary structure in the form of a tree graph with twenty nodes that cover all the vertices and nineteen of the thirty links located on the edges of the dodecahedron, leaving eleven of the thirty edges of the dodecahedron uncovered. The interaction between the nineteen links of the tree and the capsid constitute the specific interactions while the interaction of the eleven remaining edges with the capsid will be the generic contacts. Tree graphs are defined as collections of nodes connected by links such that there is one and only one path of links connecting any pair of nodes \cite{Bollobas} (see Fig.\ref{fig:dodtrees}, right). A \textit{spanning tree graph} of a polyhedron is defined as a tree graph whose nodes are located on the vertices of the polyhedron with just enough links to connect the nodes together in a tree structure \cite{Graham}. For a dodecahedron there are of the order of $10^5$ spanning trees. They represent the different possible PS configurations within the model.

\begin{figure}[htbp]
\begin{center}
\includegraphics[width=3in]{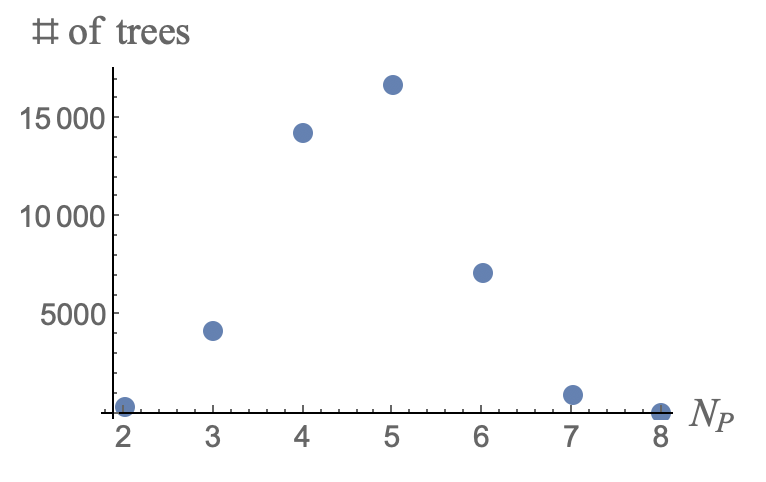}
\caption{The number of spanning trees on the dodecahedron as a function of the wrapping number $N_p$. Two spanning trees of the dodecahedron that are related by a symmetry operation of the dodecahedron are counted as the same.}
\label{fig:NP}
\end{center}
\end{figure}

Figure \ref{fig:dodtrees} (left) shows an example of a spanning tree graph of the dodecahedron. The projection of this spanning tree on the plane is shown on the right. 
Now place a pentamer on the spanning tree. Each pentamer interacts with five edges of the dodecahedron but by drawing different spanning trees one can convince oneself that a pentamer can interacts with no more than four links of a spanning tree. If the interaction of pentamer edges with the links of the spanning tree is energetically favorable -- as we will assume -- then a cluster of pentamers minimizes the interaction free energy between pentamers and spanning tree by maximizing the number of pentamers in contact with four links of the spanning tree. The figure shows that a maximum of six pentamers can be positioned in this fashion. We will say that the \textit{wrapping number} of this tree structure is $N_P=6$.  The maximum $N_P$ for a spanning tree of the dodecahedron is eight while the minimum is two. The distribution of wrapping numbers over all spanning trees is shown in Fig.\ref{fig:NP}.
The same spanning tree can be distributed over the dodecahedron in different ways with different wrapping numbers. The wrapping number is thus not a topological characteristic of the secondary structure.
  
The wrapping number measures the number of locations for an individual pentamer to be placed on a dodecahedron while making the maximum of four attractive contacts but it does not measure how many attractive contacts a newly added pentamer is able to make with pentamers that were placed earlier on the dodecahedron. The \textit{compactness} of a spanning tree is a measure of the probability that two pentamers placed on the dodecahedron are able to share an edge and the \textit{Maximum Ladder Distance} (or MLD) is a frequently used measure of the compactness of a secondary structure \cite{yoffe2008, fang}. The MLD of a spanning tree graph is defined here as the maximum number of links separating any pair of nodes. In graph theory, the ladder distance between two nodes of a tree graph is called the ``distance'' while the MLD is known as the ``diameter'' of a tree graph \cite{Bollobas}. The MLD of the tree molecule shown in Fig.\ref{fig:dodtrees} is nine. It can be demonstrated that the smallest possible MLD for a spanning tree of the dodecahedron is nine (see Appendix \ref{app:A}) while the largest possible MLD of a spanning tree is nineteen. Minimum MLD spanning trees resemble Cayley trees while maximum MLD spanning trees are \textit{Hamiltonian Paths}. The latter are walks without self-intersection that visit all vertices of a polyhedron \cite{Rudnick2005, Dykeman2013b}. In the absence of interactions, the solution radius of gyration of a branched polymeric molecule increases with the MLD as a power law \cite{gutin}. A systematic comparison between the genomic RNA molecules of RNA viruses confirms that they have significantly lower MLDs than randomized versions of the same molecules \cite{yoffe2008, fang}. It should be emphasized however that in this paper the MLD concept is applied only to the twenty link PS section.
 
\begin{figure}[htbp]
\begin{center}
\includegraphics[width=3in]{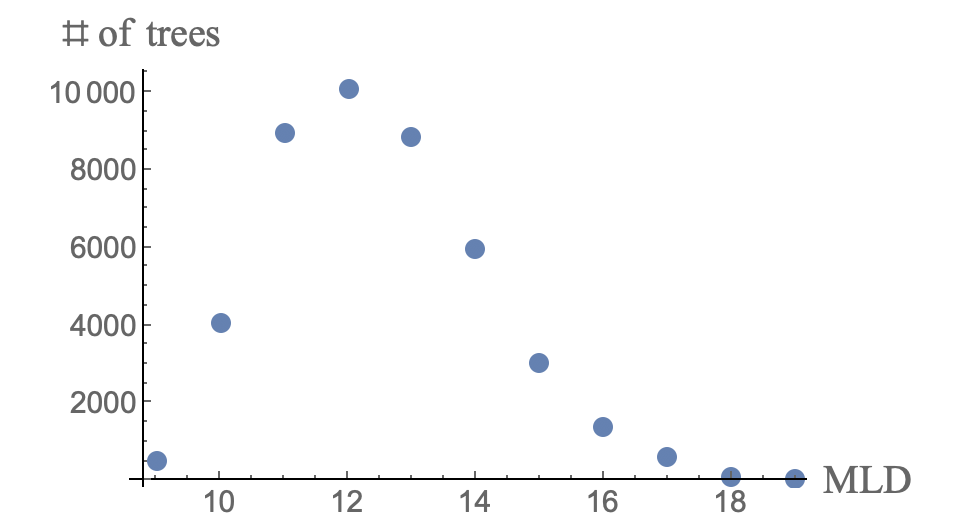}
\caption{The number of spanning trees on the dodecahedron as function of the maximum ladder distance (MLD). Two spanning trees of the dodecahedron that are related by a symmetry operation of the dodecahedron are treated as the same. }
\label{fig:dodlogplot}
\end{center}
\end{figure}

Figure \ref {fig:dodlogplot} is a plot of the number of spanning trees of the dodecahedron as a function of the MLD.
The plot has a pronounced maximum around MLD twelve. By comparison, the configurational entropy of an annealed branched polymer composed of nineteen monomers that is not constrained to be a spanning tree depends on the MLD as $19-MLD^2/19$ \cite{gutin}. It has a maximum at the smallest possible MLD. It follows that the demand that a tree molecule also is a spanning tree of a dodecahedron greatly constrains the branching statistics. 

\begin{figure}[htbp]
\begin{center}
\includegraphics[width=3in]{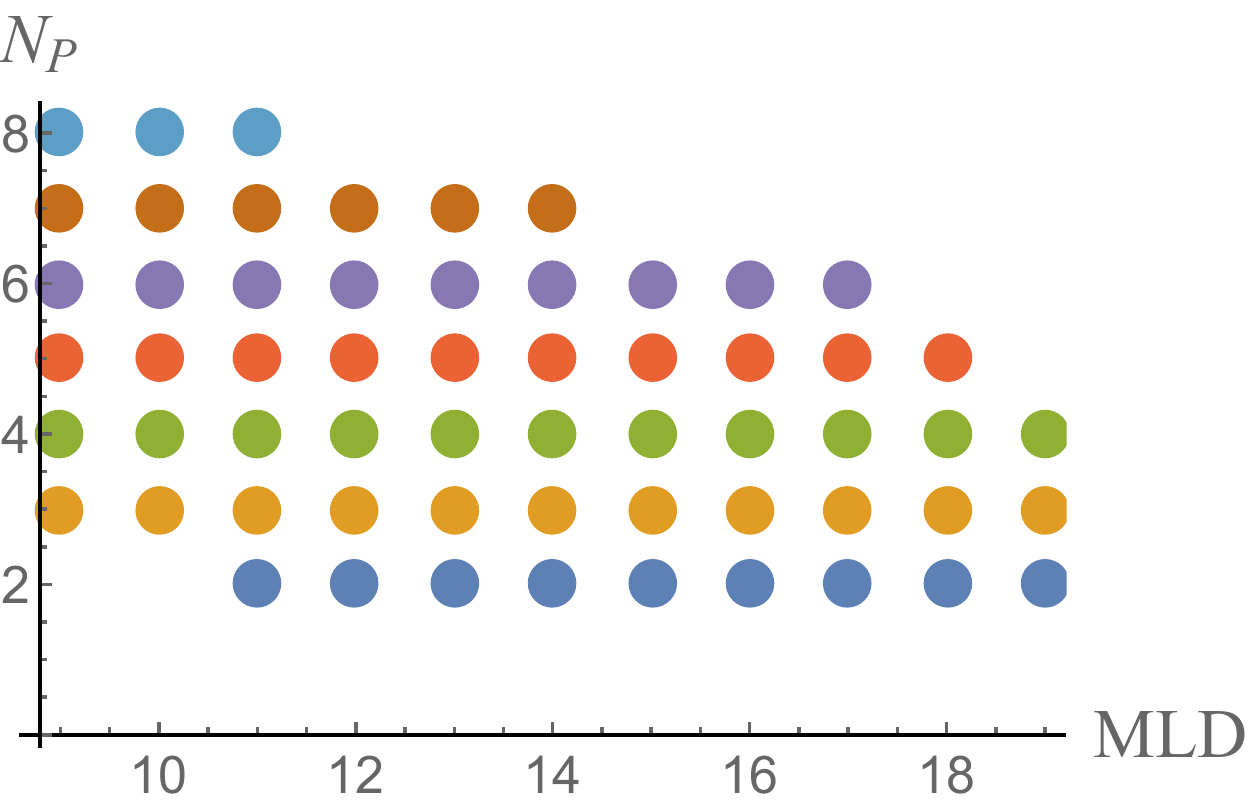}
\caption{Plot of the range of wrapping numbers ($N_P$) of the spanning tree molecules of the dodecahedron as a function of the maximum ladder distance (MLD).}
\label{fig:NP_MLD}
\end{center}
\end{figure}

The wrapping number and the maximum ladder distance are the two characteristics that we will use to classify spanning trees. Figure \ref{fig:NP_MLD} is a plot of the range of allowed wrapping numbers for given MLD. 
The two characteristics clearly are correlated. For the largest MLDs the wrapping number is restricted to be between two and four while for the smallest MLDs the wrapping number can adopt nearly its full range. The spanning trees in the upper left-hand corner are of the plot are maximally adapted for a reduced activation energy barrier while the spanning trees in the lower right-hand corner are minimally adapted.


\subsection{Assembly Energy Profiles.}

The next step is to construct the minimum energy assembly pathways and energy profiles for the Spanning Tree Model. The initial state is a spanning tree molecule folded over the edges of a mathematical dodecahedron with no pentamers. This starting state can be viewed as representing a folded or pre-condensed form of the viral ssRNA genome molecule(s) \footnote{In actuality, condensation of the RNA genome molecules takes place \textit{during} encapsidation.}. Different spanning trees are assumed to have the same folding energy prior to the binding of pentamers. Next, pentamers are placed on the dodecahedron, one after the other. The energy of a cluster of $n$ pentamers associated with spanning tree $i$ is defined as: 

\begin{equation}
\Delta E(n)/E_0 = -\left(n_1\epsilon + n_2(2\epsilon+1)+n_3 +n\mu_0\right)
\end{equation}

Here, $n_3 \leq 11$ is the number of edges shared between two pentamers that are not covered by a spanning tree link. The corresponding affinity is minus $E_0$, as it was for the Zlotnick Model. Next, $n_1$ is the number of links of the spanning tree that lie along a pentamer edge that is not shared with another pentamer. Physically, minus $\epsilon E_0$ is the the affinity of a pentamer edge with a spanning tree link. Note that $\epsilon$ is also the ratio of that same affinity with the affinity between two pentamer edges not in contact with a link of the spanning tree. Finally, $n_2$ is the number of spanning tree links that lie along a pentamer edge that \textit{is} shared with another pentamer. Interactions between edges and spanning tree links are assumed to be additive so the bond energy of such a link is $-(1+2\epsilon)E_0$. The assembly energy of a complete particle is equal to $\Delta E/E_0 = -\left(19(1+2\epsilon) +11+12 \mu\right)$ for all spanning trees, with -- as before -- $\mu_0$ the reference pentamer chemical potential. Note that the assembly energy of complete particles does not depend on the class of spanning trees in this model. 

\begin{figure}[htbp]
\begin{center}
\begin{tabular}{@{}cc@{}}
\includegraphics[width=1.75in]{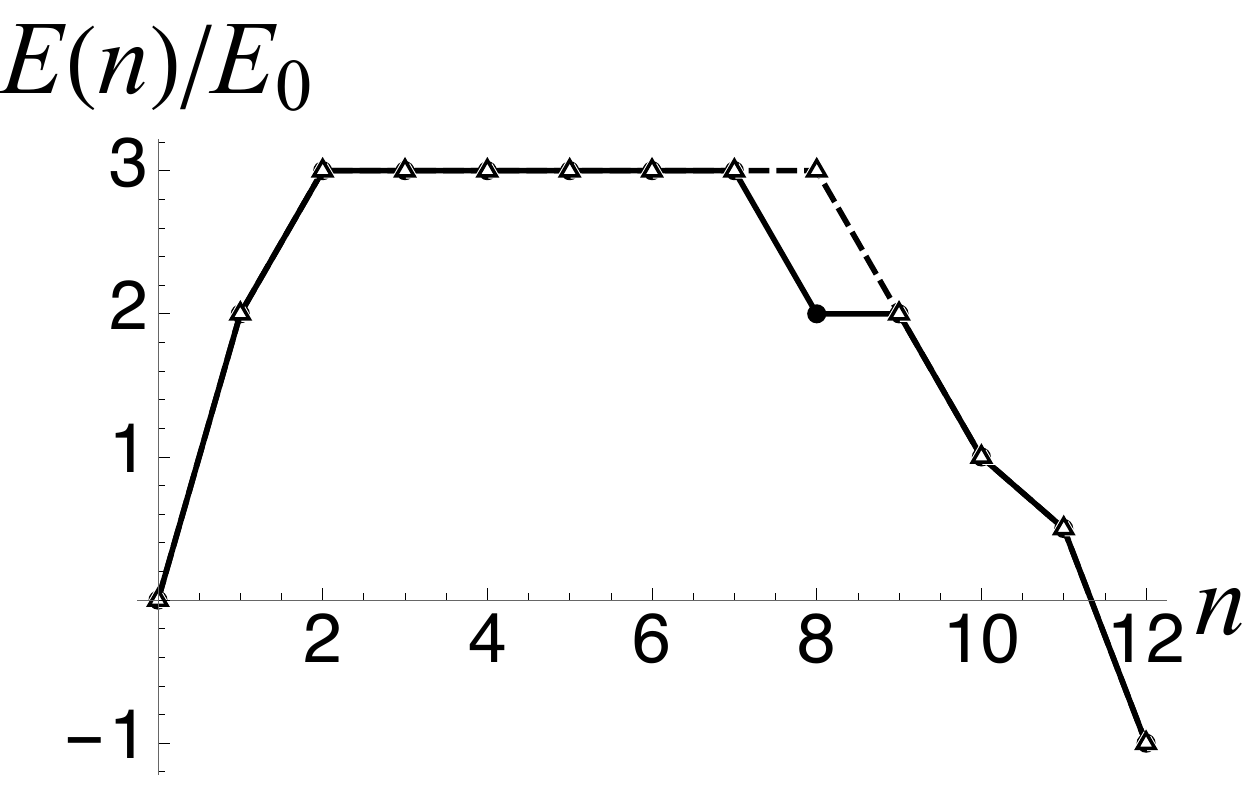} &
\includegraphics[width=1.75in]{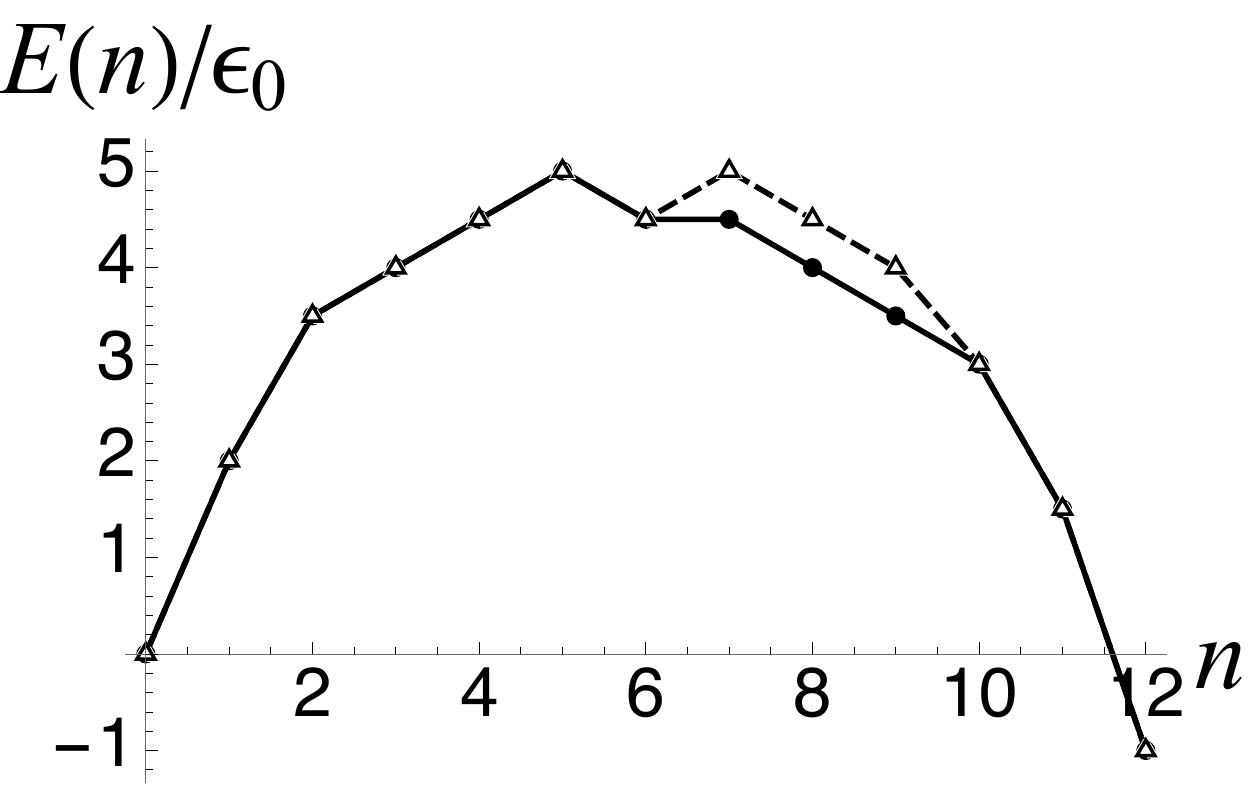} \\
\includegraphics[width=1.75in]{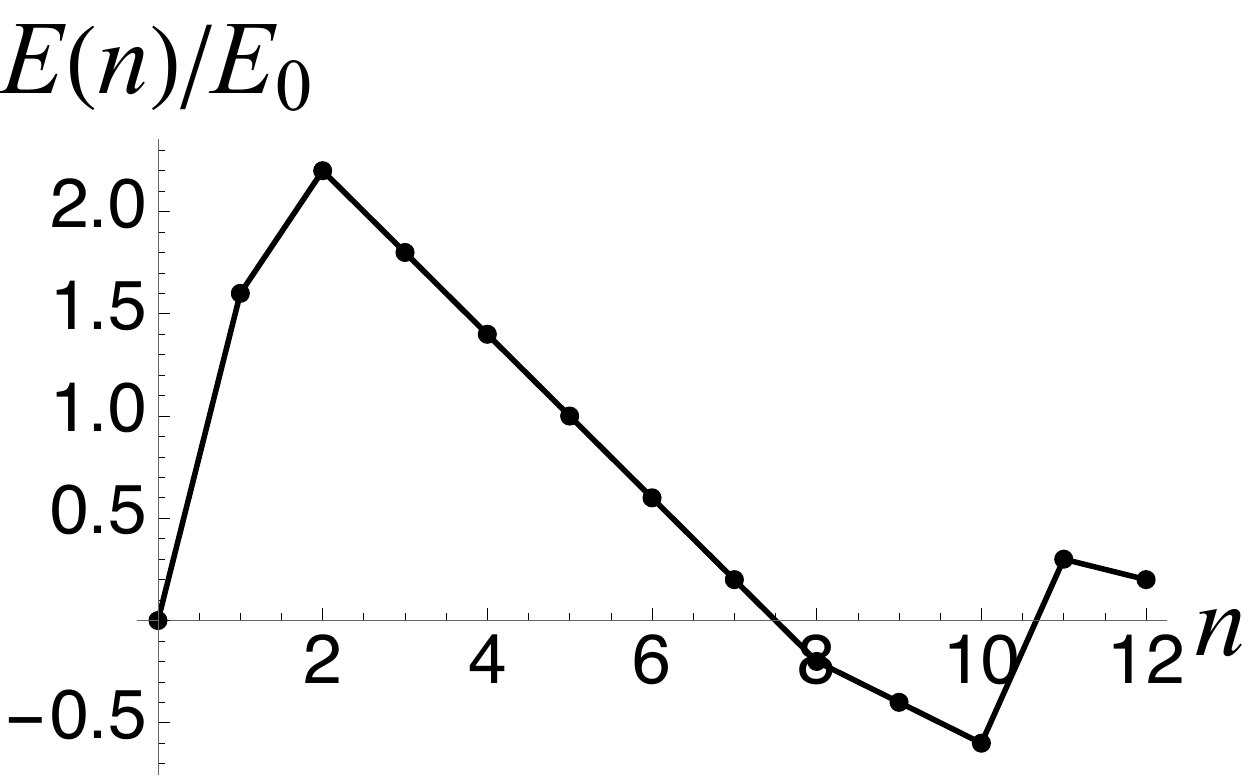} &
\includegraphics[width=1.75in]{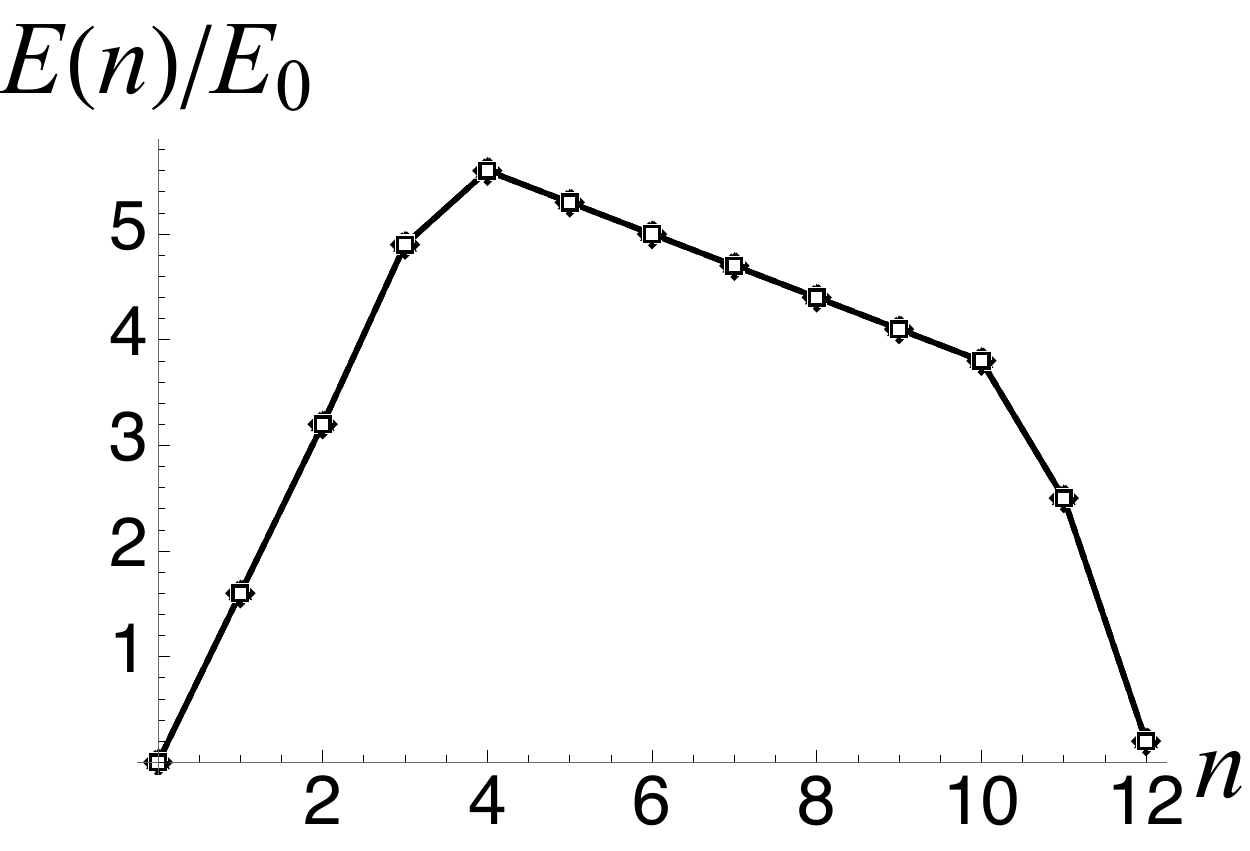}
\end{tabular}
\caption{Top: Minimum-energy assembly profiles for $N_P=8$, $MLD=9$ spanning trees (left) and for $N_P=2$, $MLD=19$ spanning trees (right). The affinity ratio is $\epsilon=0.5$ and the reference chemical potential $\mu_0=-4.0$. Energies are expressed in units of the overall scale $E_0$. Bottom: Same expect that $\epsilon=1.1$ and the reference chemical potential $\mu_0=-6.0$.}
\label{example2}
\end{center}
\end{figure}

Examples of minimum-energy assembly profiles are shown in Fig.\ref{example2}. 
The top left figure shows the assembly energy profile of $N_P=8$, $MLD=9$ spanning trees, which we will refer to as class (1). The reference chemical potential is close to that of assembly equilibrium ($\mu^*\simeq-4.083$). While there are in general many different spanning trees for a given $N_P$ and $MLD$, \textit{nearly all have the same energy profiles}. The energy profiles of the small number of exceptions is shown in the figure. Neither the $N_p$  by itself nor the $MLD$ by itself suffices as a good characteristic but their combination works quite well. This approximation leads to an important simplification: energy profiles can be satisfactorily classified by the pair of $N_p$ and $MLD$ numbers.

The right figure shows the case of $N_P=2$, $MLD=19$ Hamiltonian walk spanning trees, which we will refer to as class (2), that have no structural adjustment to bind pentamers. The activation energy is about two units of $E_0$ which means that the highly branched spanning trees of the left figure have a lower assembly activation barrier than the linear spanning trees of the right figure. Note the metastable minimum at $n=6$. Metastable intermediate states like these are familiar from experimental studies of viral assembly \cite{Parent2006, Tuma2008, Basnak2010} as well as from numerical simulations \cite{Johnston2010, Hagan2011, Baschek2012, Perlmutter2015b}. They are known as ``kinetic traps" and they may retard assembly. 

In the bottom two figures, the affinity ratio $\epsilon$ is increased to $1.1$ while the chemical potential has been reset to $-6.0$ in order to maintain the system close to assembly equilibrium. The difference between the assembly energy barriers has increased to about four units of $E_0$. This expected since increasing $\epsilon$ increases the energy contrast between pentamer bonds that are and that are not lined by an RNA link. The assembly energy profile of the $N_P=8$, $MLD=9$ spanning tree has developed a new minimum at $n=10$, which means that the minimum energy state is a particle with two missing pentamers! Breakdown of the nucleation-and-growth assembly scenario becomes frequent when $\epsilon$ is significantly larger $0.5$. Note that this did not happen for the Hamiltonian Walk.

\subsubsection*{Assembly Energy Pathways.}

\begin{figure}[htbp]
\begin{center}
\includegraphics[width=3.5in]{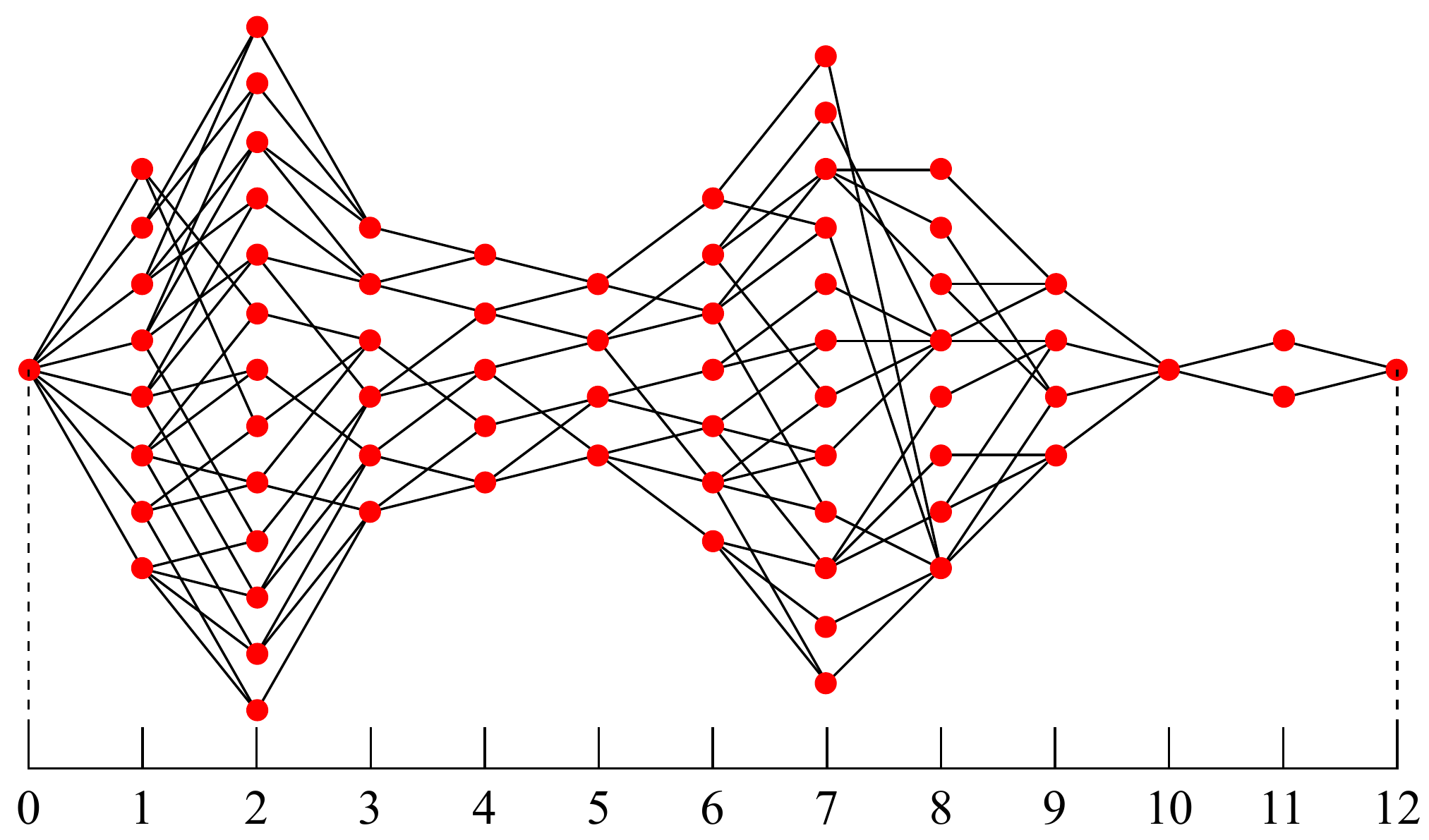}
\includegraphics[width=3.5in]{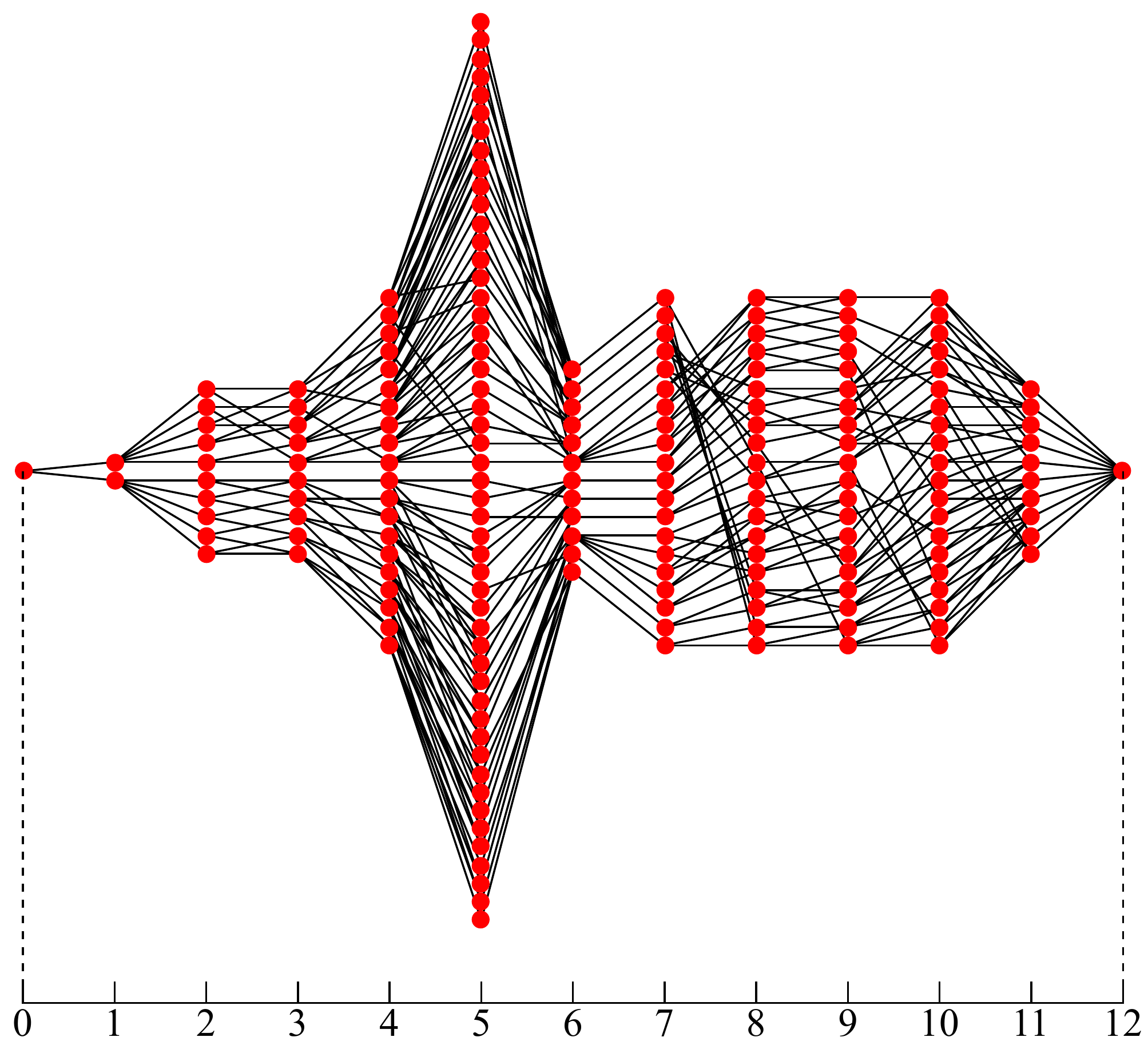}
\caption{Minimum energy assembly paths of $N_P=8$, $MLD=9$ spanning trees (top) and $N_P=2$, $MLD=19$ spanning trees (bottom). Each node (indicated by a solid dot) indicates a physically distinct intermediate structure with, from left to right, $n=0, 1, ...., 12$ pentamers. Every possible path from n=0 to n=12, including back steps, represents a possible minimum energy assembly pathway. }
\label{example3}
\end{center}
\end{figure}

Next, we explored the configuration space of minimum energy assembly pathways with results shown in Fig. \ref{example3}.  
Each node of the network stands here for a physically distinct assembly intermediate with assemblies related by a symmetry operation of the dodecahedron being treated as the same. Nodes are assigned ``coordinates" $(n,i)$ with $n=0, 1, ...., 12$ the number of pentamers of the intermediate and with $i=1, 2, ...... , m_n$ an index ranging over the distinct n-pentamer states where $m_n$ is the \textit{multiplicity} of the n-pentamer state (e.g., $m_5=4$ for the $N_P=8$, $MLD=9$ spanning tree). A black line linking two dots indicates that the two states can be interconverted by addition or removal of a pentamer. Assembly of viral particles can be viewed as a net ``current" flow from the $n=0$ source state to the $n=12$ final state along all possible paths across the network linking the initial state to the final state. Under conditions of thermodynamic equilibrium, the current across each individual link should be zero according to the principle of detailed balance. Note that the $N_P=8$, $MLD=9$ spanning spanning trees have far fewer assembly intermediates and assembly pathways than the $N_P=2$, $MLD=19$ Hamiltonian Walk spanning trees. Specifically, the multiplicity number $m(n)$ of vertical dots for given $n$ is defined as the number of distinct n-pentamer intermediates. 
\begin{figure}[htbp]
\begin{center}
\includegraphics[width=3.5in]{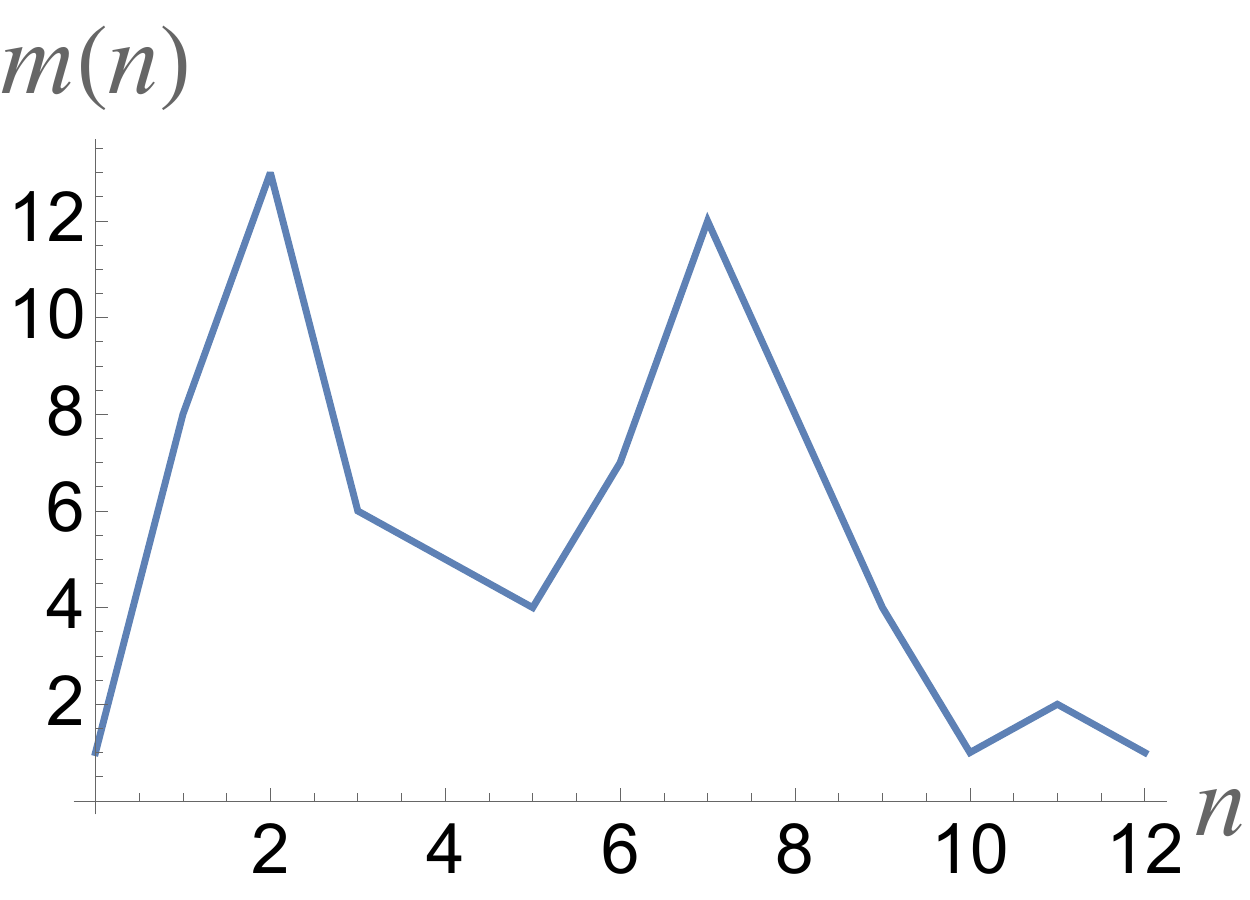}
\includegraphics[width=3.5in]{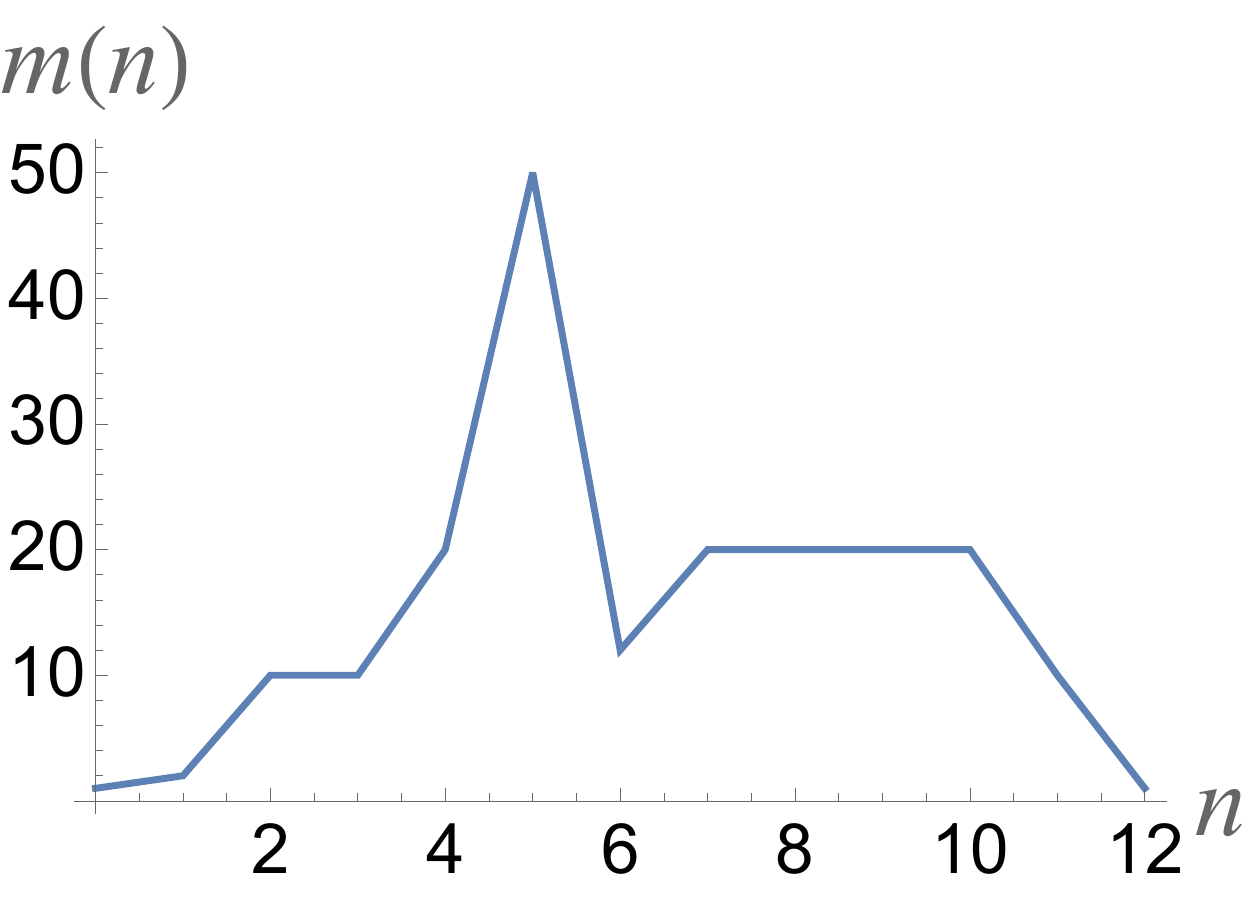}
\caption{Multiplicities $m(n)$ of the $N_P=8$, $MLD=9$ spanning trees (top) and the $N_P=2$, $MLD=19$ spanning trees (bottom). }
\label{example4}
\end{center}
\end{figure}
The multiplicity of the n=5 assembly intermediates of the $N_P=2$, $MLD=19$ spanning trees is about $10^2$ times larger than that of the $N_P=8$, $MLD=9$ spanning trees.
 
Assuming that the assembly energy profiles of spanning trees with the same $N_P$ and $MLD$ are all the same allows us to define a ``low temperature" Boltzmann Distribution for all RNA molecules with a particular $N_P$ and $MLD$:
\begin{equation}
P_{B}(n)\propto\exp-\Delta F(n)
\label{BD}
\end{equation}
where $\Delta F(n)= \beta\Delta E(n) - \ln m(n) - n\ln c_f(eq)$. The second term includes the entropic free energy associated with the multiplicity $m(n)$ of an n-pentamer assembly and the third term the correction to the pentamer solution chemical potential for the general case that the equilibrium concentration of free pentamers $c_f(eq)$ differs from the reference concentration (which is our unit of concentration). With ``low-temperature" we mean here that the energy scale $E_0$ of the assembly energy profile is sufficiently large compared to the thermal energy that we only need to include n-pentamer assembly intermediates that minimize the assembly energy for given n. In Appendix B we discuss the equilibrium phase behavior obtained from this Boltzmann Distribution. It turns out to be typical of that of self-assembling systems in general with a critical pentamer concentration (``CAC") below which viral particles do not form and above which the particle concentration increases linearly with the pentamer concentration. The concentration of packaged particles obeys the Law of Mass Action of chemical thermodynamics. 

\subsubsection*{Structural Transitions.}

\begin{figure}[htbp]
\begin{center}
\includegraphics[width=3.5in]{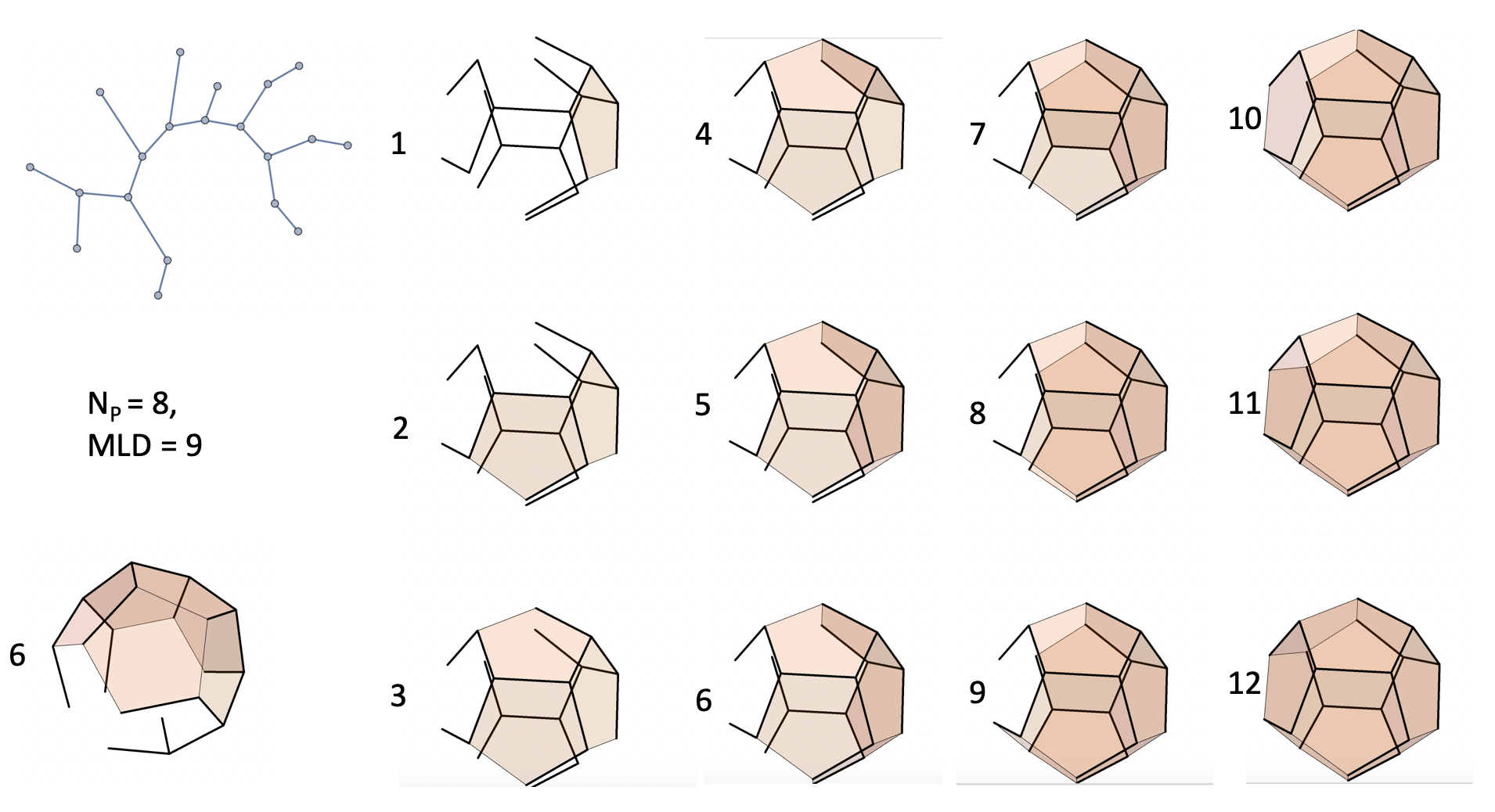}
\caption{Assembly pathway for a $N_P=8$, $MLD=9$ spanning tree spanning tree for the case of small $\epsilon$. The first five pentamers can be placed on sites that maximize both the number of pentamer-pentamer contacts and pentamer-spanning tree link contacts. The sixth pentamer, shown separately with a different perspective, makes only two spanning tree link contacts, which allows it to still have three pentamer-pentamer contacts. Note the similarity with Fig.2.}.\label{HS}
\end{center}
\end{figure}

\begin{figure}[htbp]
\begin{center}
\includegraphics[width=2.0in]{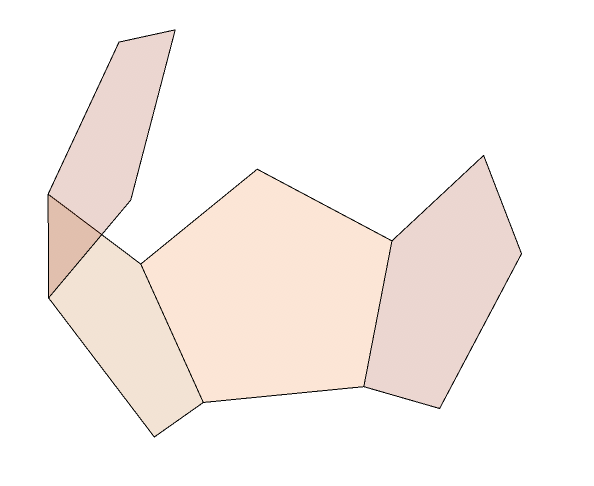}
\caption{The minimum energy $n=4$ assembly state of a class (2) molecule for $\epsilon=1.2$.}
\label{fig:F}
\end{center}
\end{figure}

For $\epsilon$ small compared to one, the pentamers are most often placed on minimum energy sites where they make the maximum number of edge-to-edge contacts with previously placed pentamers. The resulting assembly intermediates are compact pentamer clusters, similar or the same to the ones shown in Fig.2 for the Zlotnick Model. An example is shown in Fig.\ref{HS} .
On the other hand, for $\epsilon$ large compared to one the first $N_p$ pentamers typically are placed on maximum wrapping sites. This indicates the possibility for a \textit{structure transition} of assembly intermediates as a function of $\epsilon$. For example, for small $\epsilon$ six-pentamer clusters have five-fold symmetry with one central pentamer sharing its five edges with five other pentamers that each share three edges with their neighbors (see the n=6 state of Fig.11). On the other hand, for large $\epsilon$ a minimum energy $n=6$ cluster of class (1) $N_P=8$, $MLD=9$ spanning trees has the six pentamers placed on the six available maximum wrapping sites of the $N_P=6$ spanning tree (see Fig.4). By moving just one pentamer, the two structures can be transformed into one another. This transition takes place at $\epsilon=-1$. For class (2), with $N_p=2$, the transition is more dramatic. For small $\epsilon$, the $n=4$ pentamer cluster is a compact structure with a two-fold symmetry axis, the same as the n=4 structure shown in Fig.2. On the other hand, the $n=4$ minimum energy structure for large $\epsilon$ shown in Fig.\ref{fig:F} is completely different.

This linear arrangement of pentamers has an interesting feature. Allow tree links to rotate freely around the nodes of the tree and allow pentamers to swivel freely around shared edges. The pentamers of the empty-capsid partial assemblies of Fig.2 would -- for $n>2$ -- not be able to move with respect to each other without breaking pentamer-pentamer bonds. The empty-capsid partial assemblies can be said to be mechanically rigid. The same holds  for the $n=6$ structure of Fig.4 and other small $\epsilon$ partial assemblies. However, this is not the case for the four-pentamer structure shown in Fig.\ref{fig:F}: if this structure were allowed to fluctuate freely, then the four pentamers could freely swivel along the three shared edges. Structural transitions of this type as a function of $\epsilon$ become more common for larger values of the MLD.

\section{Kinetics and Packaging Competition.}
In order to construct the kinetics we start by characterizing graphs of the assembly pathways of spanning tree with given $N_P$ and MLD in terms of an \textit{adjacency matrix} $A_{n}^{i,j}$. The adjacency matrix equals one if a link connects node $(n,i)$ to node $(n+1,j)$ and zero if there is no link. Next, define for each node $(n,i)$ of the network a time-dependent occupation probability $P_{i,n}(t)$. The kinetics is assumed to be Markovian with the probabilities evolving in time according to the Master Equation \cite{vanKampen}:
\begin{equation}
\begin{split}
&\frac{dP_{i,n}(t)}{dt}=\\&\sum_j\{A_{n-1}^{j,i}W_{n-1,n}P_{j,n-1}(t)+A_{n}^{i,j}W_{n+1,n}P_{j,n+1}(t)\} \\&-P_{i,n}(t)\sum_j\{A_{n-1}^{j,i}W_{n,n-1}+A_{n}^{i,j}W_{n,n+1}\}
\label{eq:ME}
\end{split}
\end{equation}
Here, $W_{n,n+1}$ is the on-rate for the transition of an assembly of $n$ pentamers to one with size $n+1$ by the addition of a pentamer while $W_{n,n-1}$ is the off-rate at which a pentamer is removed from an assembly of size $n$. Physically, the assumption of Markovian kinetics amounts to assuming a pentamer-by-pentamer assembly scenario (an alternative scenario will be discussed in Section IV). Next, we will assume a simplified diffusion-limited chemical kinetics \footnote{See Supplementary Material (2)} in which the addition or removal of a pentamer to an assembly of size $n$ is treated as a bimolecular reaction with an on-rate that has the form of a kinetic Monte-Carlo algorithm:
\begin{equation}
 W_{n,n+1}=\lambda c_f(t)
 \begin{cases}
 e^{-\Delta\Delta E_{n,n+1}}\quad\thickspace \text{if}\quad \Delta E(n+1) > \Delta E(n)\\
1\qquad\quad\qquad\quad \text{if} \quad \Delta E(n+1) <  \Delta E(n)
 \end{cases}
 \end{equation}
The concentration $c_f(t)$ of free pentamers is in general time-dependent, and different from the reference concentration, because assembly of capsids reduces the concentration of free pentamers. The on-rates are thus time-dependent as well. Next $\Delta\Delta E_{n,n+1} = \Delta E(n+1)-\Delta E(n)$ is the energy cost of adding a pentamer while $\lambda$ is a base rate that depends on molecular quantities such as diffusion coefficients and reaction radii but not on the pentamer and RNA concentrations. The inverse of $\lambda$ is the fundamental time-scale of the kinetics and, in the following, time will be expressed in units of $1/\lambda$. If $\Delta\Delta E_{n,n+1}$ is negative then the on-rate is equal to this base rate. If $\Delta\Delta E_{n,n+1}$ is positive then the base rate is reduced by the Arrhenius factor  $e^{-\Delta\Delta E_{n,n+1}}$. 

The off-rate entries $W_{n+1,n}$ are determined first by the condition that in the long-time limit the occupation probabilities must approach the equilibrium Boltzmann distribution. This imposes the condition of detailed balance $\frac{W_{n,n+1}}{W_{n+1,n}}|_{t\to\infty}=\frac{P_{B}(n+1)}{P_{B}(n)}=c_f(eq) e^{\Delta\Delta E_{n,n+1}}$. We also demand, on physical grounds, that the off-rates for the release of a Gag protein from a cluster should be independent of the concentration of free pentamers. Both conditions are satisfied by imposing.
\begin{equation}
\frac{W_{n,n+1}}{W_{n+1,n}}=c_f(t) e^{\Delta\Delta E_{n,n+1}}
\end{equation}

The ratio of the free pentamer concentration $c_f(t)$ over the total time-independent pentamer concentration $c_0$ is determined by pentamer number conservation:
\begin{equation}
c_f(t)/c_0 = 1 - (D/12)\sum_{n=0}^{12}\left(\sum_{i=1}^{m(n)} n P_{i,n}(t)\right)
\label{eq:gammadef1}
\end{equation}
Here, $D\equiv12 r_t/ c_0$, with $r_t$ the total RNA concentration, is the RNA to protein \textit{mixing ratio}. If $D=1$ then there are exactly enough pentamers to encapsidate all spanning trees, which corresponds to the \textit{stoichiometric ratio}. Because all occupation probabilities enter in the relation for $c_f(t)$ that itself enters in all thirteen equation, the rate equations form in fact a coupled set of non-linear differential equations. In the following sections we will solve these equations by numerical integration using Mathematica. 

\subsection{Time-Scales}

\begin{figure}[htbp]
\begin{center}
\includegraphics[width=3.5in]{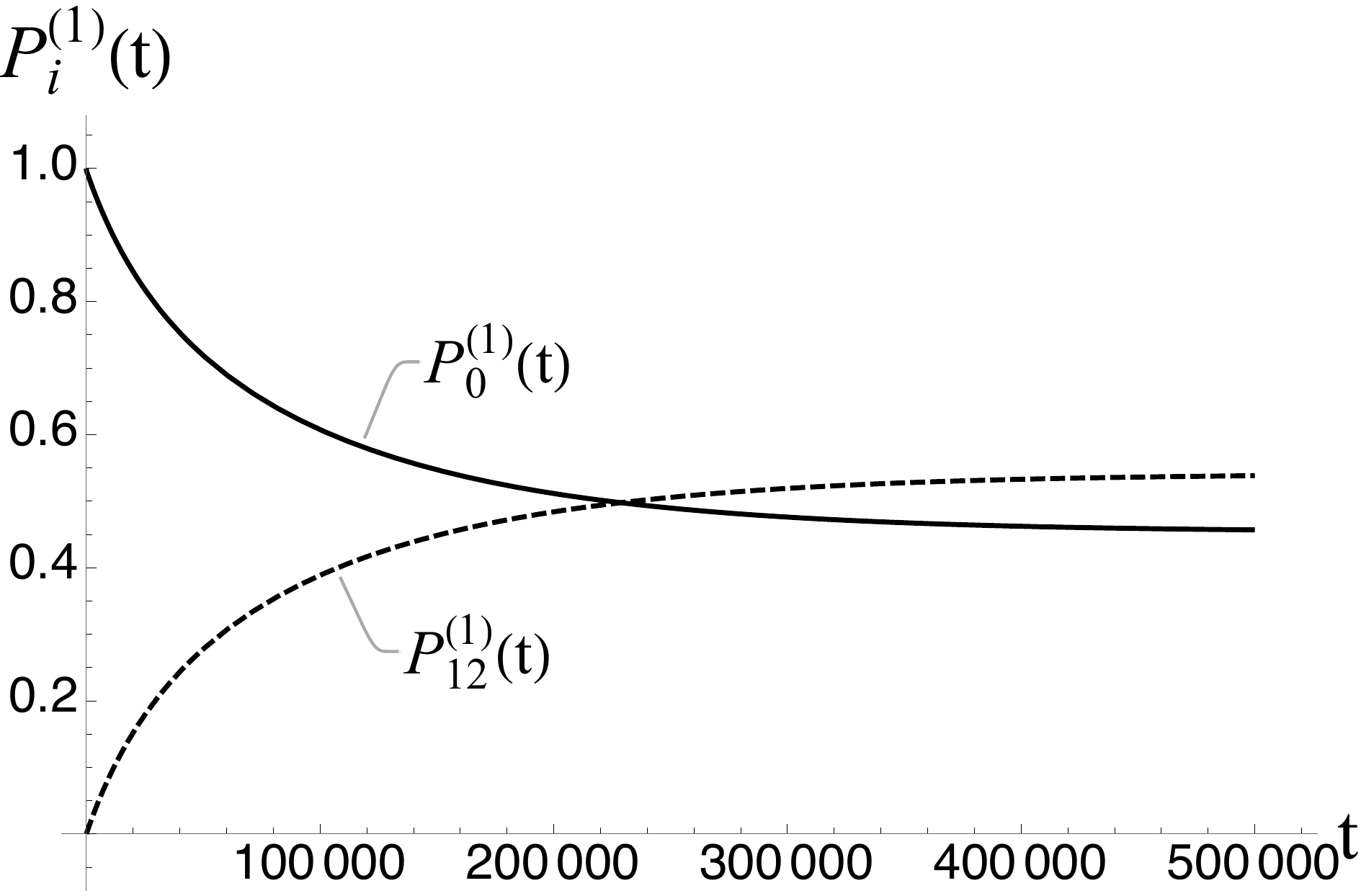}
\includegraphics[width=3.5in]{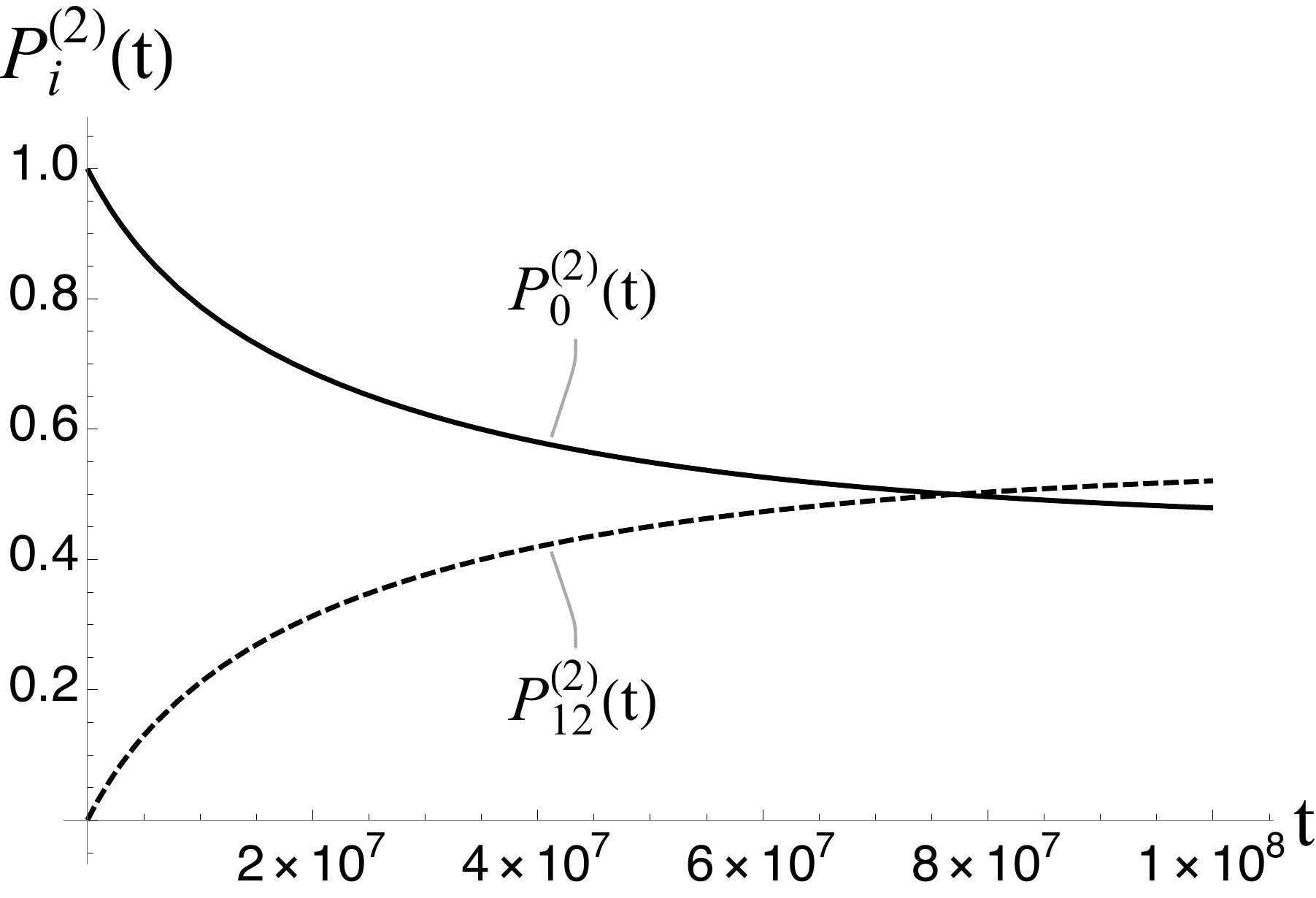}
\caption{Top: Packaging kinetics of MLD = 9 and $N_p = 8$ spanning trees. Parameter values are $E_0=4 k_bT$ for the energy scale, $\epsilon=0.5$, $c_0=1$, $D=0.5$ and $\mu_0=-4$. Bottom: Packaging kinetics of MLD = 19, $N_p = 2$ spanning trees with the same parameters.}
\label{RS1}
\end{center}
\end{figure}

Figure \ref{RS1} shows the packaging kinetics of the two classes of RNA molecules.  For the overall energy scale we used a value $E_0=4 k_bT$ close to that of the pentamer-pentamer affinity of the Zlotnick Model for empty capsids. For the ratio $\epsilon$ between RNA/pentamer to pentamer/pentamer interaction we set $\epsilon=0.5$. For this value, the nucleation and growth scenario is still applicable (see Fig.\ref{example2}). Next, the total pentamer concentration was set to $c_0=1$ and the mixing ratio to $D=0.5$. The second condition means that there are twice as many pentamers as would be necessary to package all RNA molecules. Finally, we set the reference chemical potential at $\mu_0=-4$, close to the assembly equilibrium chemical potential.
  
Both systems approach In the late time limit the thermal equilibrium state with roughly the same fraction of RNA molecules being packaged (about sixty six percent). This reflects the fact that the two classes of molecules have the same assembly energy. The remaining difference is due to the fact the entropy term of the assembly free energy is not the same for the two classes. The reason that a significant fraction of RNA molecules are not being packaged, despite the fact that there are twice as many pentamers as needed to package the RNA molecules, reflects the fact that the chemical potential is close to assembly equilibrium. That means the fully assembled and fully disassembled state have comparable statistical weights in the Boltzmann distribution. While the time dependence of the occupation probability has the same shape, the relaxation times are quite different. Roughly $10^5$ time units for MLD=9 and $N_p=8$ spanning trees and $10^7$ time units for MLD=19, $N_p=2$ spanning trees. This difference is consistent with the fact that the assembly activation barrier is about $2E_0$ larger (so about $8 k_bT$) for the MLD=19, $N_p=2$ spanning trees.

 In order to compute relaxation times, one first completes the definition of the rate matrices by introducing the diagonal entries $W_{n,n}=-\sum_{m\neq n}W(m,n)$. The resulting matrix $W_{m,n}$ now has column elements adding to zero. Using this completed transition matrix, the master equation can be rewritten in the form of the matrix equation $\frac{d\bf{P}}{dt}=\bf{W P}$. This looks like a linear matrix equation but because the concentration of free pentamers is self-consistently dependent on all occupation probabilities through Eq.\ref{eq:gammadef1}, the matrix $\bf{W}$ depends on the occupation probabilities so this is not the case. However in the long-time limit when the system is close to thermal equilibrium, one can replace the occupation probabilities in Eq.\ref{eq:gammadef1} by the equilibrium Boltzmann probabilities in Eq.\ref{eq:gammadef1} to obtain the spectrum of relaxation rates for a system in thermal equilibrium. The equation $\frac{d\bf{P}}{dt}=\bf{W P}$ can be solved by standard matrix diagonalization methods. The eigenvalues are the late time decay rates of the various modes that correspond to the eigenvectors. The lowest relaxation rate, which determines the approach to final equilibrium, denoted by $t_r$,  is the inverse of the smallest eigenvalue of $W_{m,n}$. This gives $t_r\simeq3.26\times 10^5$ for the MLD=9, $N_p=8$ spanning trees and $t_r\simeq3.4\times 10^7$ for the MLD=19, $N_p=2$ spanning trees. 

\begin{figure}[htbp]
\begin{center}
\includegraphics[width=3.0in]{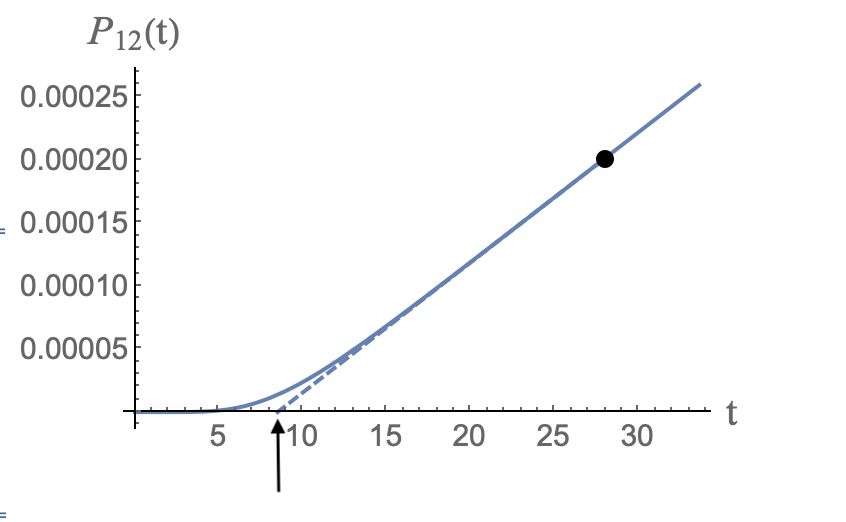}
\caption{Definition of the delay time as the intersection of the tangent to the assembly curve $P_{12}(t)$ with maximum slope with the time axis. $MLD=9$, $N_p=8$, $\epsilon=0.5$, $c_0=1$, $D=0.5$ and $\mu_0=-4$. }
\label{Lag}
\end{center}
\end{figure}

This thermalization time can be compared with the early-time assembly \textit{delay time} $t_d$. This is the time lag between the establishment of solution assembly conditions and the first appearance of assembled viral particles. Measured delay times for the assembly of empty capsids are in the range of minutes \cite{Prevelige1993, Casini2004, medrano}. We obtain $t_d$ from the intersection of the tangent to $P_{12}(t)$ at the point of maximum slope with the horizontal axis (see Fig.\ref{Lag}). 

For the case of the MLD=9 and $N_p=8$ class of spanning trees, this gives about $8.5$ time units, so four to five orders of magnitude smaller than the thermalization time. Other classes have comparable delay times. Comparing with experimentally measured delay times for the assembly of empty capsids indicates that the time unit $1/\lambda$ is in the range of 1-10 seconds. The thermalization time under conditions of assembly equilibrium would then be in the range of \textit{two hundred hours} for MLD=9, $N_p=8$ spanning trees and two orders of magnitude longer for the MLD=9, $N_p=8$ spanning trees.  However, in vitro assembly experiments are carried out on supersaturated solutions. When the reference pentamer chemical potential $\mu_0$ is raised to $-3.6$ then the thermalization time is reduced to about $8.3\times10^3$ while the delay time remains about the same. The thermalization time would then be in the range of hours.

\subsection{Packaging Competition.}

The kinetic equations can be extended to the case of packaging competition between RNA molecules belonging to two different classes, say (1) and (2), that are competing for pentamers. If the solution contains equal amounts of the two RNA molecules then the two occupation probabilities $P^{(1,2)}_{i,n}(t)$ obey a Master Equation for the respective thirteen occupation probabilities. These two sets of equations are coupled because the same free pentamer concentration appears in both sets of equation. This free pentamer condition is determined by the condition of pentamer number conservation, which now takes the form
\begin{equation}
c_f(t)/c_0 = 1 - (D/24)\sum_{n=0}^{12}\left(\sum_{i=1}^{m_n^{(1)}} n P^{(1)}_{i,n}(t)+\sum_{i=1}^{m_n^{(2)}} n P^{(2)}_{i,n}(t)\right)
\label{eq:gammadef2}
\end{equation}

\begin{figure}[htbp]
\begin{center}
\includegraphics[width=3.5in]{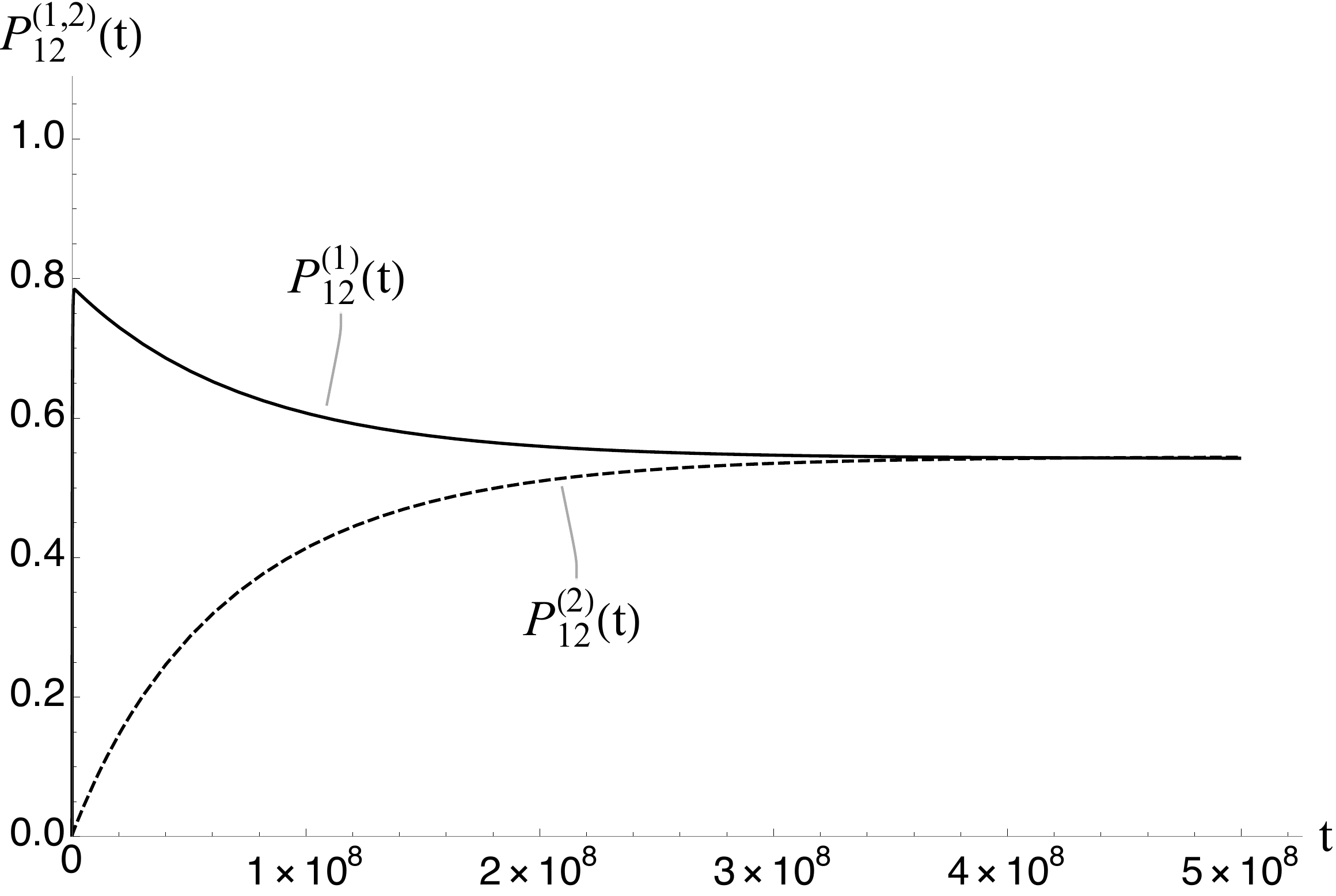}
\includegraphics[width=3.5in]{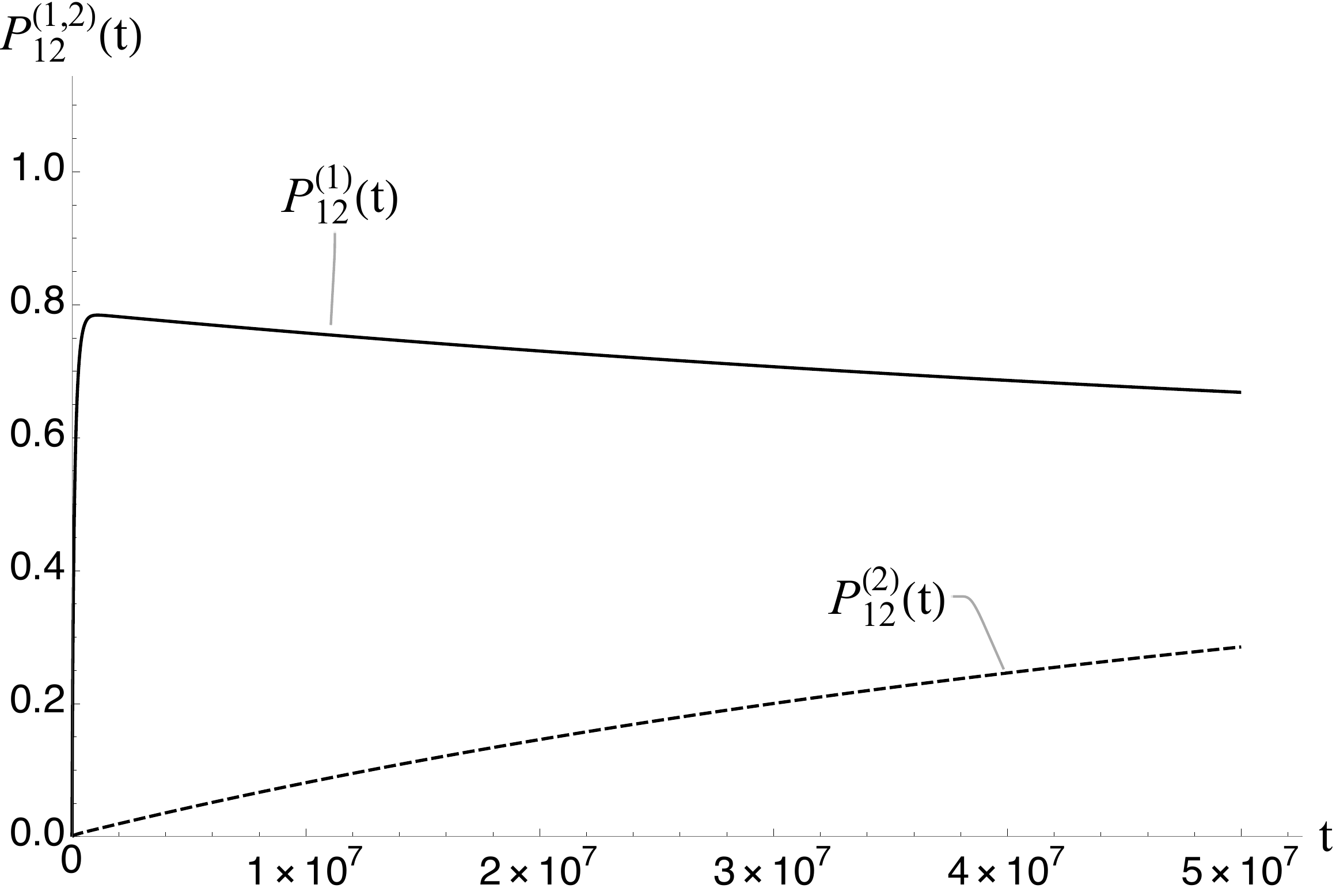}
\includegraphics[width=3.4in]{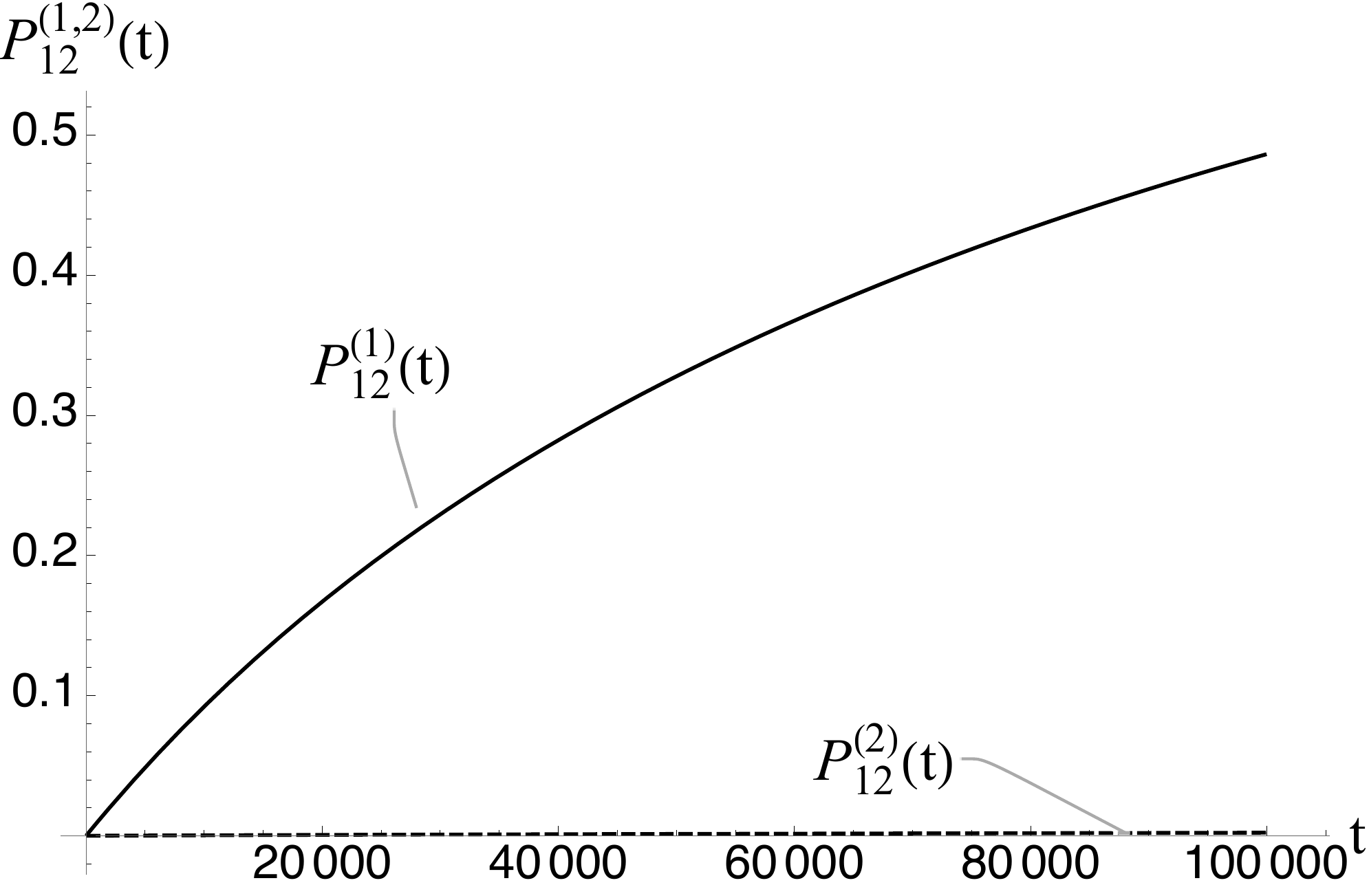}
\caption{Packaging competition between MLD=9, $N_p=8$ spanning trees and MLD=19, $N_p=2$ spanning trees with the same parameters as the previous figure. Top: time scale $10^8$ units; Middle: time scale $10^7$; Bottom: time scale $10^4$ units. }
\label{RS2}
\end{center}
\end{figure}

In Fig.\ref{RS2}, we show the outcome of a packaging competition experiment with the same total amount of RNA molecules and pentamers as before but now with half of the RNA molecules being MLD=9 and $N_p=8$ spanning trees and the other half MLD=19, $N_p=2$ spanning trees.

The packaging of MLD=9, $N_p=8$ spanning trees dominates on all time scales less than about $10^7$ time units. About eighty percent of these spanning trees are packaged around $10^7$ time units while only ten percent the MLD=19, $N_p=2$ spanning trees are packaged. The packaged fraction of MLD=9, $N_p=8$ spanning trees slowly decreases on time scales of the order of $10^7-10^8$ time units,  which means that packaged MLD=9, $N_p=8$ spanning trees are gradually \textit{disassembling}. Disassembly of viral particles is indeed essential for reaching a state of complete thermal equilibrium under conditions of packaging competition. The fraction of packaged MLD=19, $N_p=2$ spanning trees increases correspondingly. Apparently, pentamers that are being freed up by disassembly of MLD=9, $N_p=8$ spanning trees feed assembly of the MLD=19, $N_p=2$ spanning trees. The bottom figure shows what happens on times scales of the order of the thermal relaxation time of the  MLD=19, $N_p=2$ trees. When the system approaches thermal equilibrium, the packaging fractions of the two classes are nearly the same and not far from the equilibrium value found before in the absence of competition. 

\begin{figure}[htbp]
\begin{center}
\includegraphics[width=3.5in]{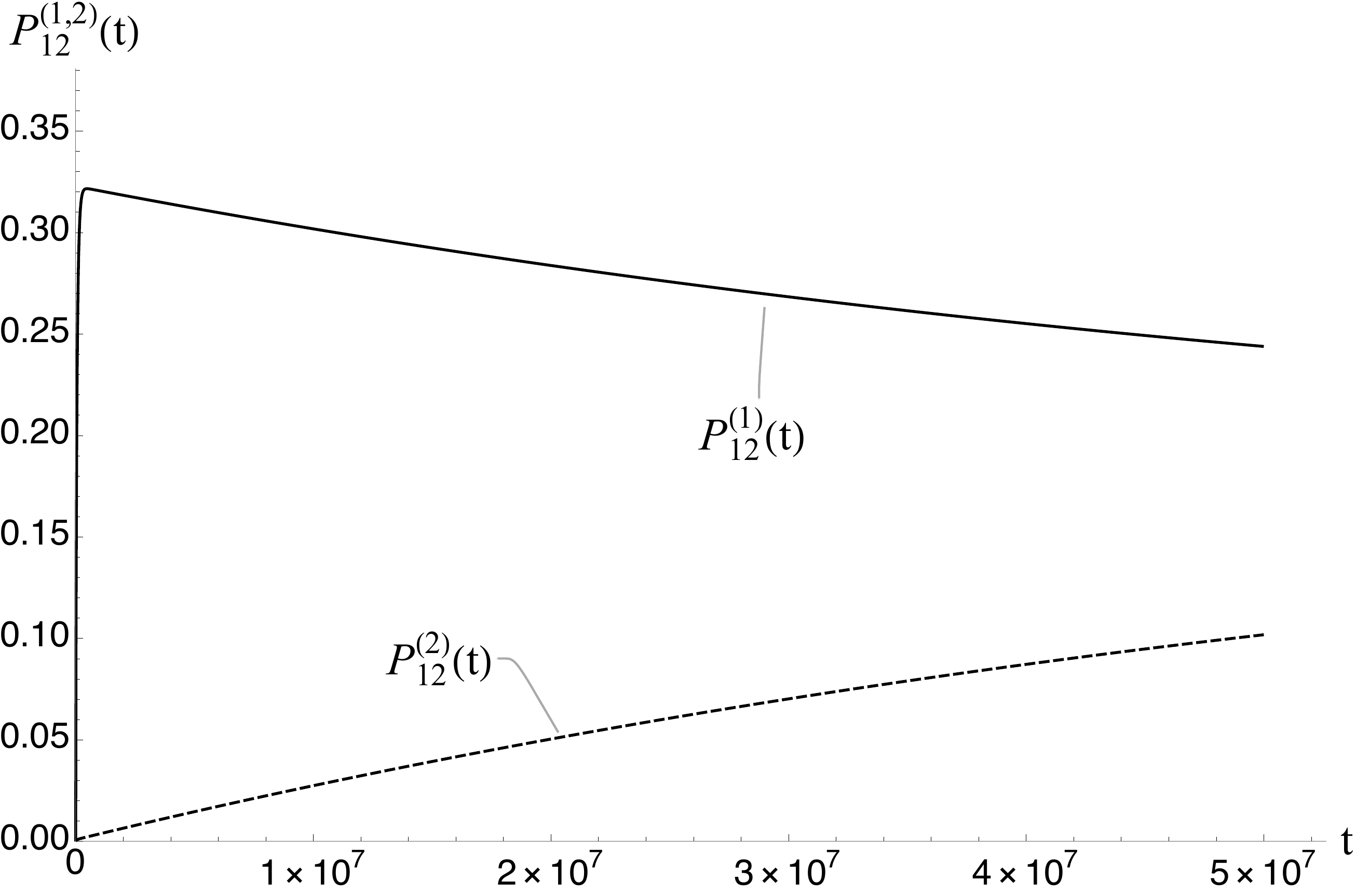}
\caption{Packaging competition between MLD=9, $N_p=8$ spanning trees and MLD=19, $N_p=2$ spanning trees with the same parameters as the previous figure except that the mixing ratio is increased to $D=2$. }
\label{RS3}
\end{center}
\end{figure}

We could now explore how this kinetic form of RNA selection is influenced by changes in the control parameters. One key quantity turns out to be the mixing ratio $D$. For the $D=0.5$ value used so far, there is a significant excess of pentamers. On the other hand, for $D=2$ there would be enough pentamers to package half of all RNA molecules. We reasoned that if the early packaging of MLD=9 and $N_p=8$ particles would deplete the available pentamers then this would ``starve" the subsequent assembly of MLD=19, $N_p=2$ spanning trees, thereby extending the time interval over which the packaging of MLD=9, $N_p=8$ spanning trees dominates. As shown in Fig.\ref{RS3}, this does not quite happen.
The fraction of packaged MLD=19, $N_p=2$ spanning trees at $5\times 10^7$ time units does decrease, from about $0.25$ to about $0.12$, but the fraction of packaged MLD=9, $N_p=8$ spanning trees \textit{also} decreases, from about $0.7$ to about $0.27$. The increase of the mixing ratio increased only marginally the relative fraction of packaged MLD=9, $N_p=8$ spanning trees. 

\begin{figure}[htbp]
\begin{center}
\includegraphics[width=3in]{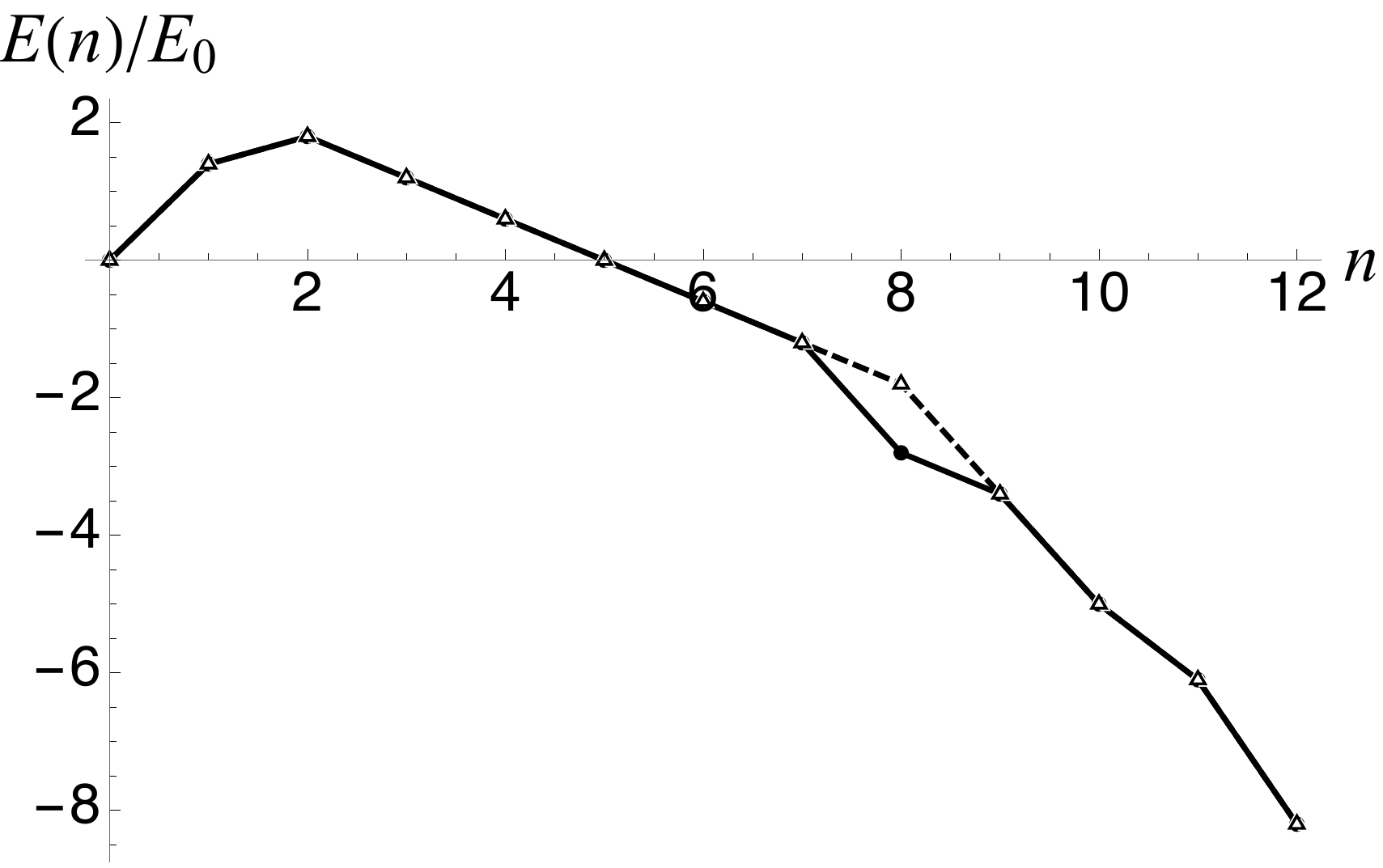}
\includegraphics[width=3in]{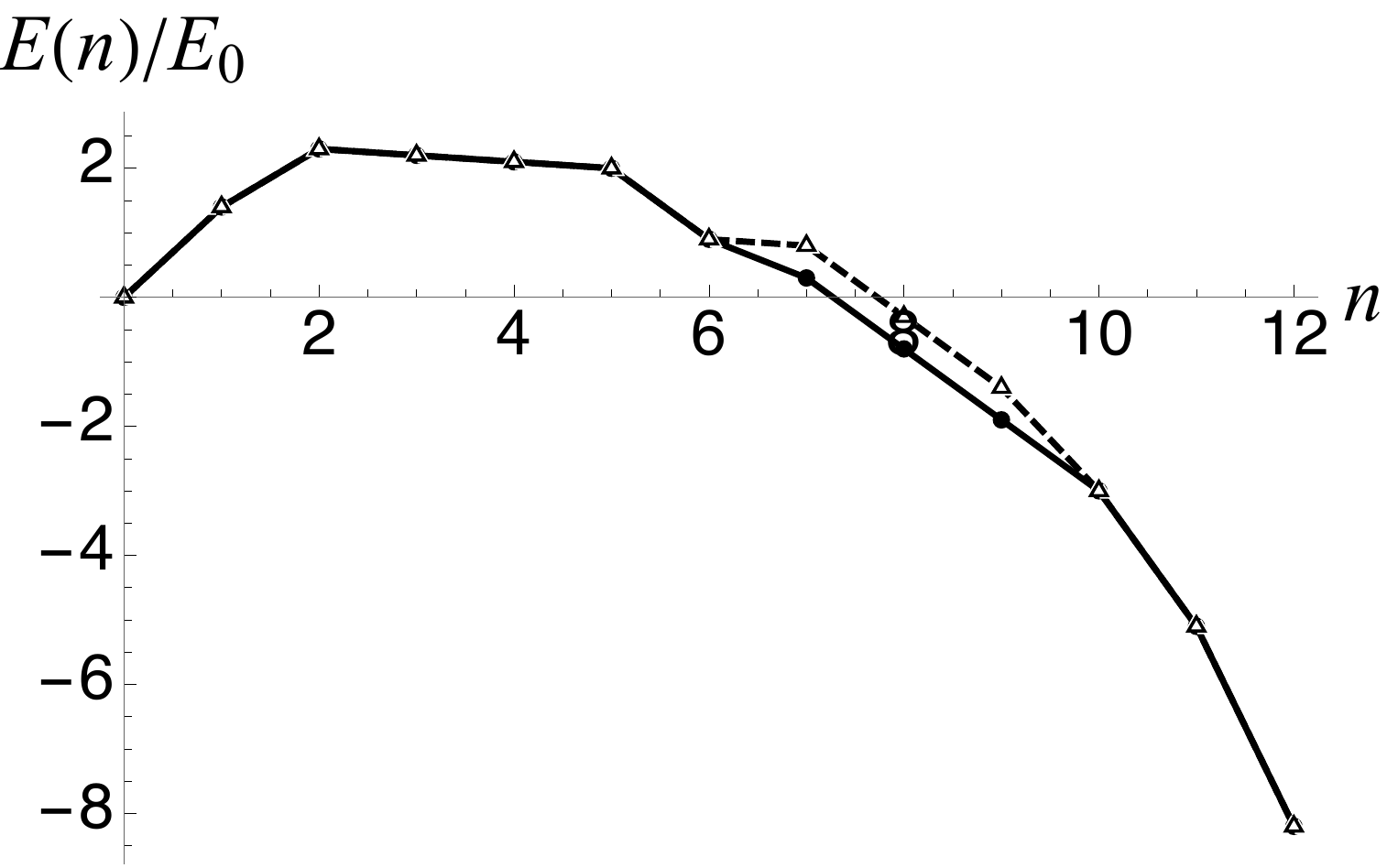}
\includegraphics[width=3.5in]{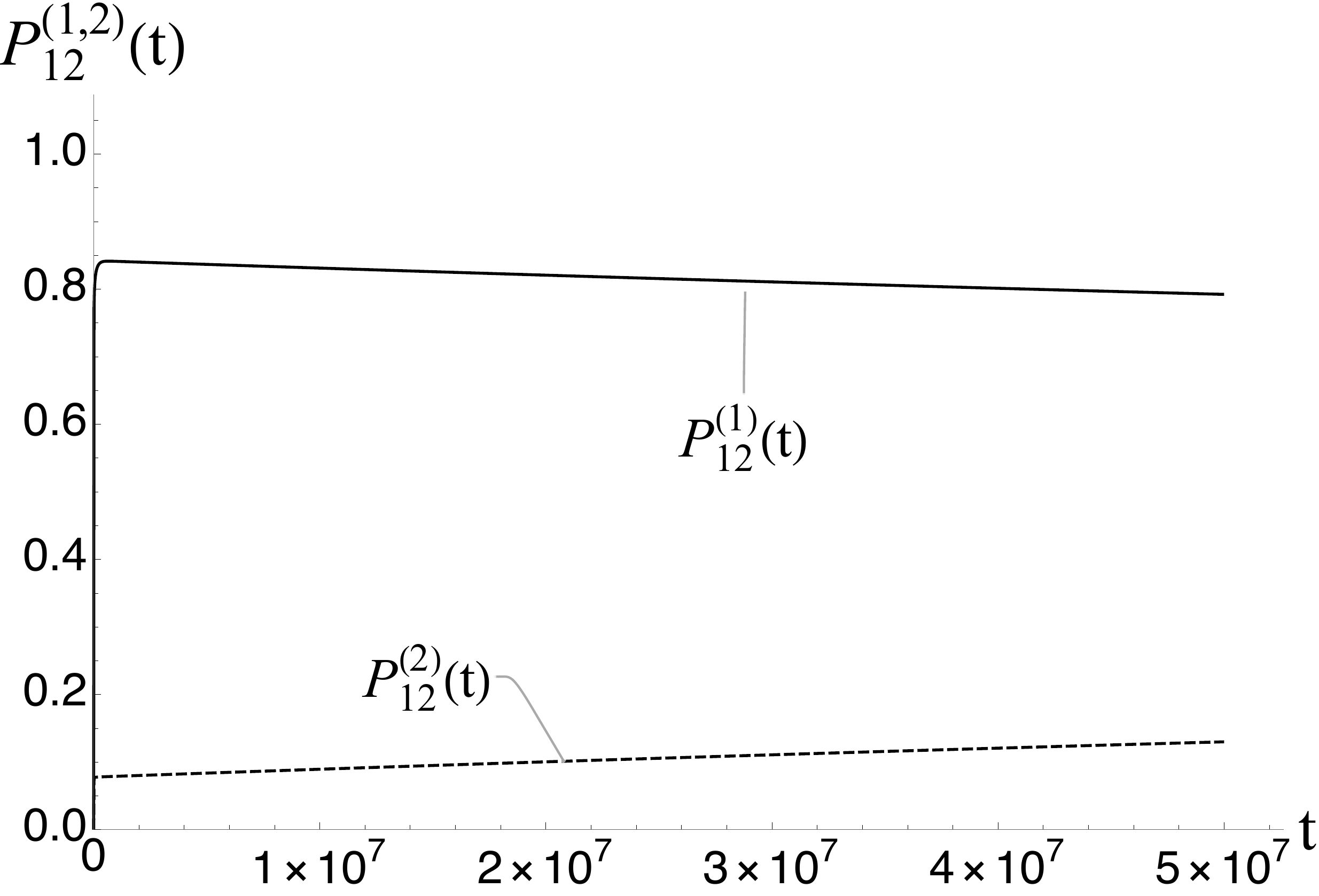}
\caption{Top and middle figures: effect on the assembly energy profiles of reducing the reference chemical potential from $\mu_0=-4$ near assembly equilibrium to $\mu_0=-3.4$. Bottom: Packaging competition between MLD=9, $N_p=8$ spanning trees and MLD=19, $N_p=2$ spanning trees with the same parameters as the previous figure except that $\mu_0=-3.4$.}
\label{RS4}
\end{center}
\end{figure}

We then examined the effect of supersaturation on packaging competition by reducing the reference chemical potential from $\mu_0=-4.0$ to $\mu_0=-3.4$. The results are shown in Fig.\ref{RS4}.
The top and middle figures show the minimum energy assembly profiles of the two classes. The assembly activation energy has decreased by about $3 E_0$ for the first class and by about $5 E_0$ for the second class and both are now about $2 E_0$. Because the two activation energy barriers are similar, one might expect that the kinetic selectivity will be weakened by supersaturation. The bottom figure shows that the opposite is true: supersaturation greatly increases packaging selectivity!. On a time scale of about $5\times 10^7$ time units, the fraction of MLD=9, $N_p=8$ spanning trees has increased from about $0.32$ to about $0.81$ while the fraction of MLD=19, $N_p=2$ spanning trees remains below $0.2$ over the measurement period. An important factor is that the disassembly of complete particles, which is essential for thermal equilibration, has greatly slowed down. The reason is that supersaturation increases the energy barrier for the disassembly of completed particles to about $10 E_0$ for both classes (see Fig.\ref{RS4}). During the early assembly of the MLD=9, $N_p=8$ spanning trees, there were few free pentamers left in solution since for $D=2$ there are just enough pentamers to package the MLD=9, $N_p=8$ spanning trees. Then, in the absence of much disassembly of the MLD=9, $N_p=8$ particles, the MLD=19, $N_p=2$ spanning trees now are starved of pentamers. This ``monopoly mechanism" can never work for the case of assembly equilibrium since there is always a significant fraction of free pentamers at assembly equilibrium. An additional factor is that the height of the activation energy barrier does not fully characterize the rate of barrier crossing. For the MLD=9, $N_p=8$ spanning trees, the n=2 state does function as a true transition state since there is a substantial energy drop for the n=3 state and larger states as well as for the n=1 state (see Fig.\ref{RS4} top). However, for the MLD=19, $N_p=2$ trees, the n=2 state is not a transition state: the whole interval between n=2 and n=5 has roughly the same energy (see Fig.\ref{RS4}, middle). The probability that a cluster of size n=3, 4, and n=5 can ``fall back" to the n=1 state remains quite large. Kinetic selection in favor of the MLD=9, $N_p=8$ spanning trees remains active.

\section{Two-Stage Assembly.}

A protein-by-protein assembly scenario, as implicitly embodied in the Master Equation, is not the only option. Numerical simulations of coarse-grained model systems \cite{Perlmutter2014,Perlmutter2015} reported that a collective assembly process, called the  \textit{en-masse} scenario, is possible as well. It was encountered for higher values of protein-genome affinity as compared to the protein-protein affinity. In this scenario, the genome molecules initially are in a swollen state due to electrostatic self-repulsion, and free of capsid proteins. When a genome molecule starts to capture capsid proteins, because of the generic electrostatic affinity, a disordered nucleo-protein condensate forms. As the number of captured proteins increases, the condensate shrinks because the negative charges of the genome molecule are increasingly being neutralized by positive capsid protein charges. As the shrinkage continues, the attractive protein-protein interactions become more important. Spatial ordering of the captured capsid proteins produces the viral particle. An order-disorder transition of this type in which rotational symmetry is broken can be described by Landau theory and applied to viral assembly \cite{rudnick2019}. Experiments on the encapsidation of linear \textit{double-stranded} genome molecules by capsid proteins \cite{van2020} have been interpreted according to this en-massed scenario while related modes of collective assembly have been proposed as well \cite{zandi2020}. Can we us our model to test the degree of RNA selectivity within the en-masse scenario? Because in the Spanning Tree Model the genome molecule is a compact structure right from the start, it cannot capture the transition from a swollen to a collapsed state. However a fascinating in-vitro assembly experiment that mimics the en-masse scenario was carried out in ref.\cite{Garmann2014}. During a first stage, the pH level was set at a level at which the protein/RNA affinity was large with respect to the protein/protein affinity. Disordered and incomplete assemblies were observed to form. In the second stage, the pH level was set at a level such that the protein/protein affinity was increased with respect to the protein/RNA affinity. The disordered condensates of the first stage transformed into virus-like particles. 

\begin{figure}[htbp]
\begin{center}
\includegraphics[width=3.5in]{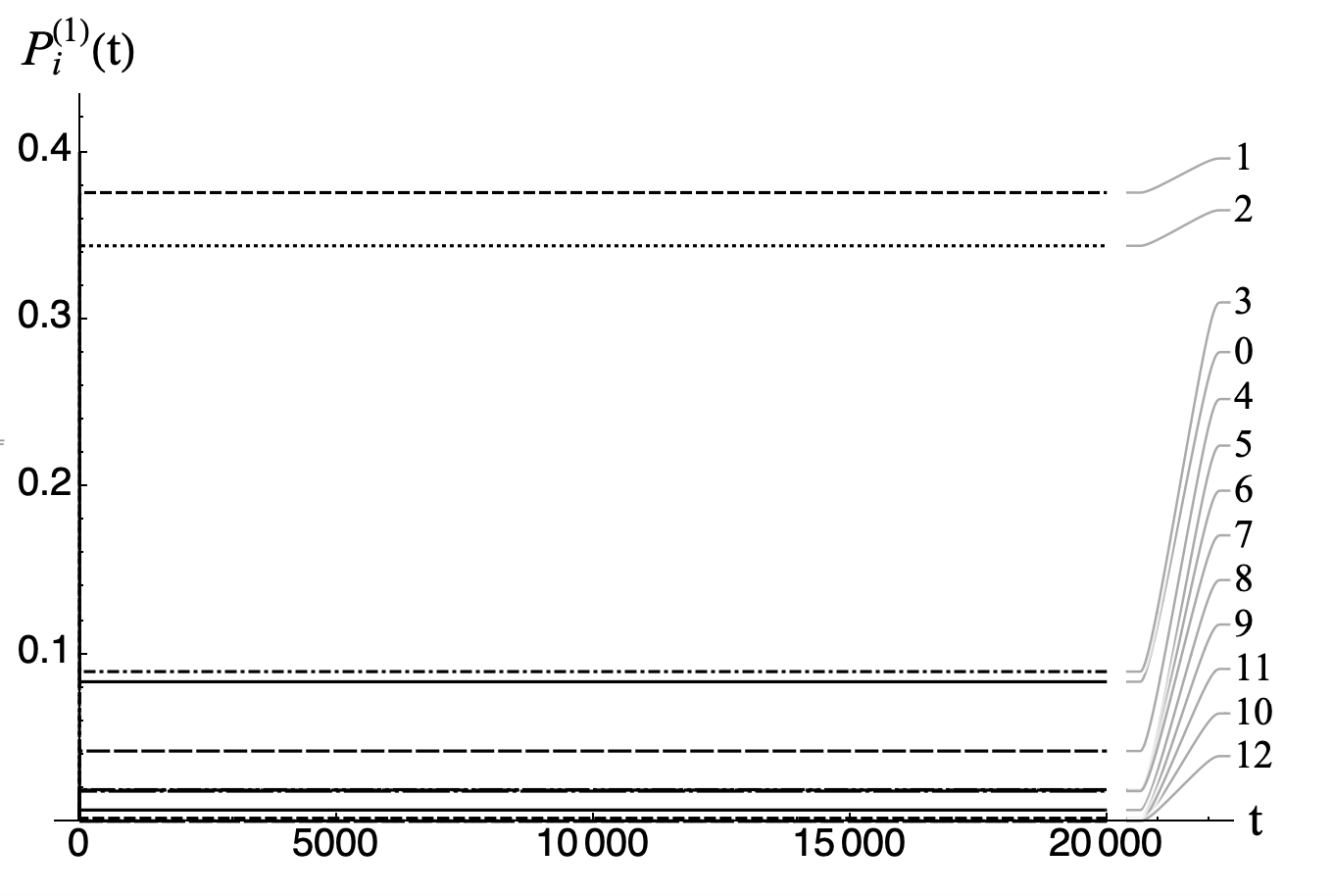}
\includegraphics[width=3.5in]{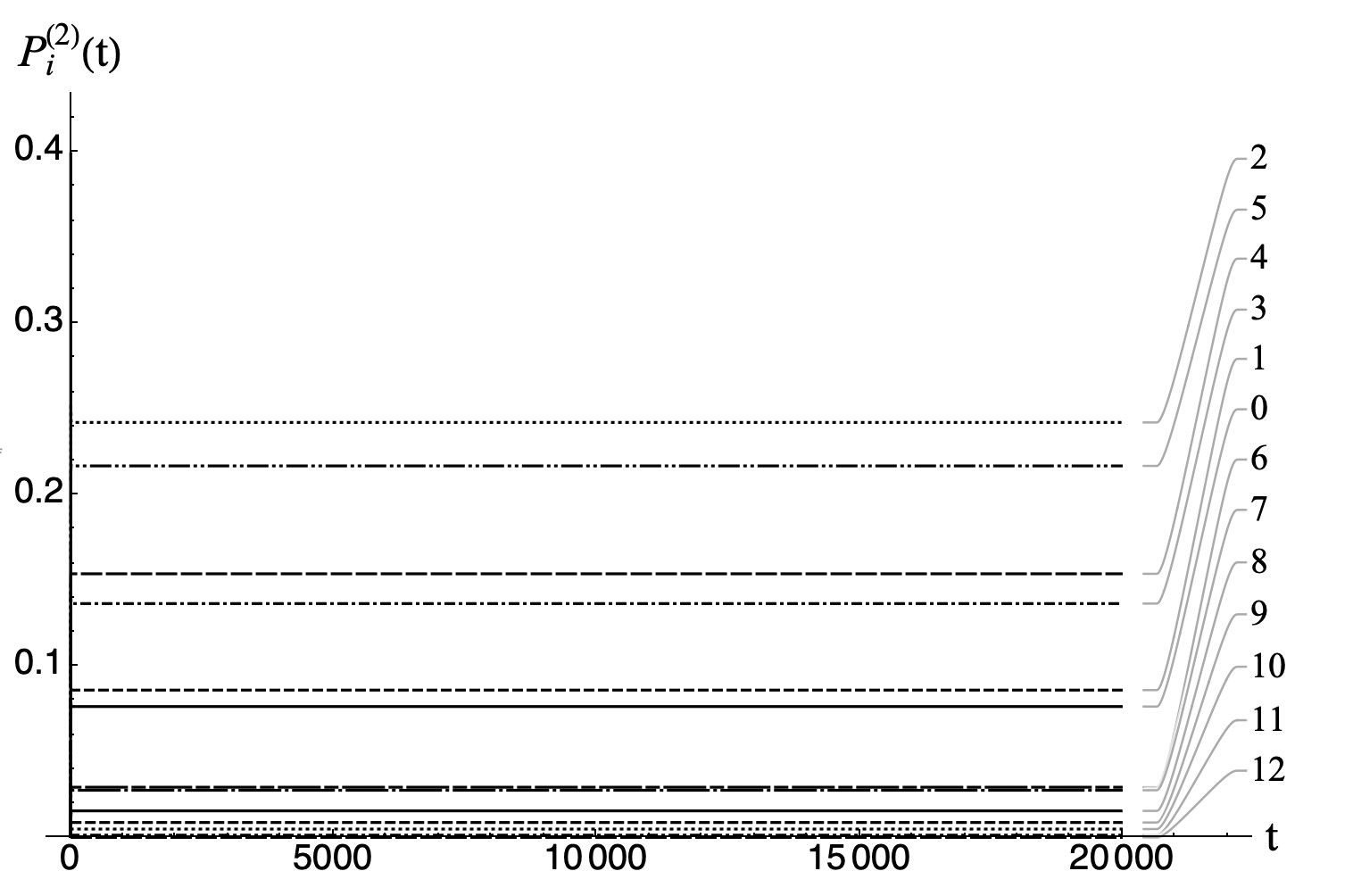}
\caption{First-stage assembly competition between MLD=9, $N_p=8$ and MLD=19, $N_p=2$ molecules. The two figures show the first-stage occupation probabilities of the MLD=9, $N_p=8$ molecules and of the MLD=19, $N_p=2$ molecules. During this stage the assembly energy profile was set to zero.}
\label{first-stage}
\end{center}
\end{figure}

\begin{figure}[htbp]
\begin{center}
\includegraphics[width=3.5in]{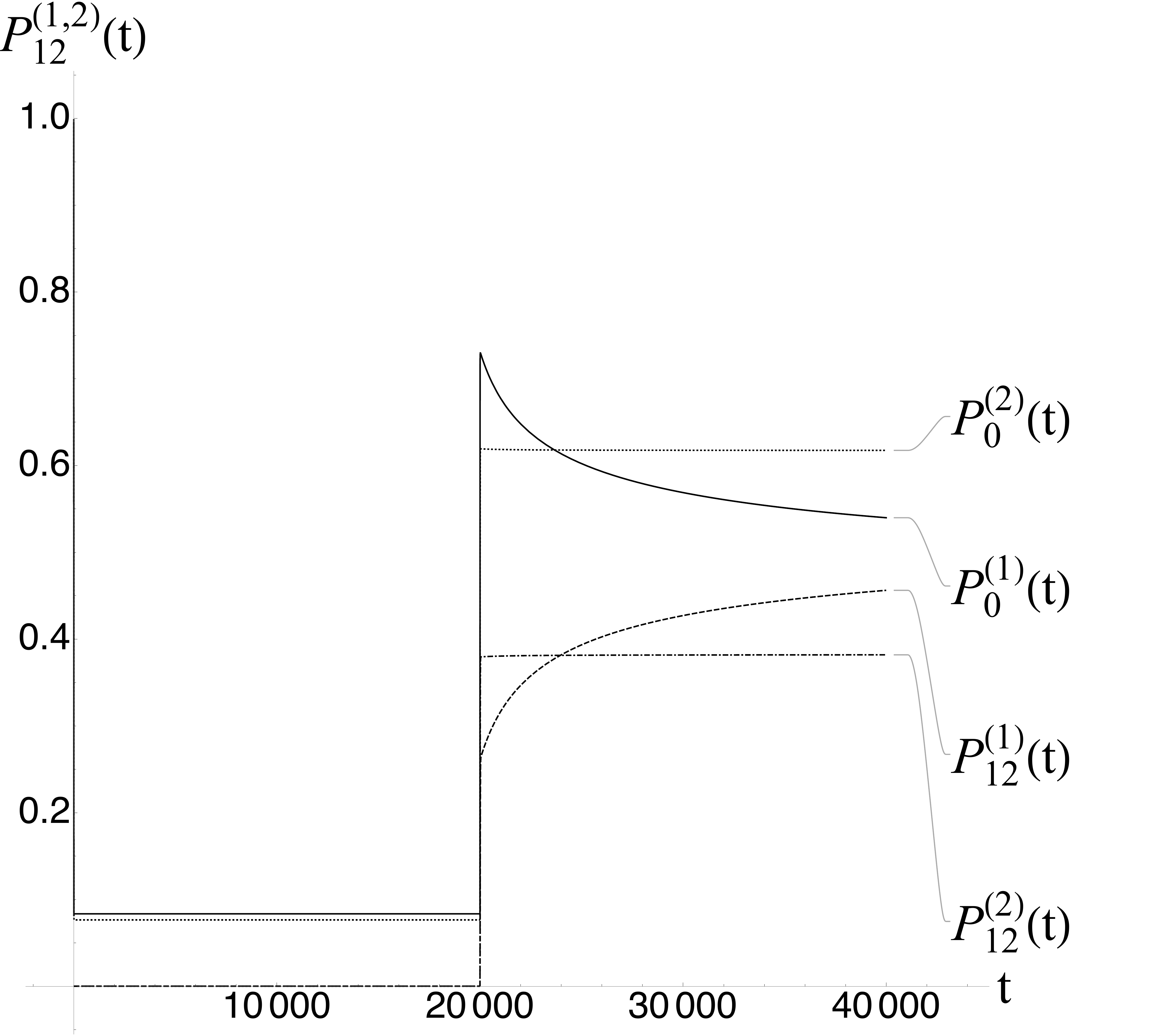}
\caption{First and second-stage occupation probabilities. At $t=20,000$, the energy parameters were reset to the values of Fig.\ref{RS4}. The occupation probabilities for assembly intermediates are negligible throughout almost all of the second state and are not shown for either stage. The fraction of packaged particles containing MLD=19, $N_p=2$ molecules is time independent while the fraction of packaged particles containing MLD=9, $N_p=8$ molecules increases moderately with time.}
\label{two_stage}
\end{center}
\end{figure}

The selectivity for such a two-stage assembly scenario \textit{can} be examined within our model namely by performing two subsequent assembly calculations. During the first stage, the assembly energy profile is set to zero in order to mimic a state in which the affinity between the capsid proteins can be neglected. Equilibration is very rapid, producing a polydisperse, disordered state of incomplete aggregates. This disordered state is then used as the \textit{initial state} for a second assembly calculation, but now with the same energy parameters as those of Fig.\ref{RS4}. Recall that in that case single-stage assembly was highly selective. Figure \ref{first-stage} show the first-stage occupation probabilities of the same two classes as before (see Fig.\ref{RS4}). 
The first-stage occupation probabilities are time-independent and consistent with the equilibrium Boltzmann distribution. The MLD=9, $N_p=8$ trees are mostly occupied by one or two pentamers but a significant fraction of the MLD=9, $N_p=8$ trees carry five, four, or three pentamers. This is a consequence of the larger multiplicities for this class, as shown in Fig.10. Figure \ref{two_stage} shows what happens when at $t=20,000$ the energy parameters are reset to the values of Fig.\ref{RS4}. 

Instantly, there is a complete reorganization. The intermediate-sized clusters produced during the first stage disappear, leaving behind fully assembled particles plus free RNA molecules of both classes. At $t=20,000$ the fraction of packaged MLD=19, $N_p=2$ molecules exceeds the fraction of MLD=9, $N_p=8$ molecules. This is a consequence of the fact that before the reorganization there were more clusters on the MLD=19, $N_p=2$ molecules with n=4 and 5, as a consequence of the larger multiplicities of these states (see Fig.10). After the reorganization, these clusters slide down the slope of the assembly curve of Fig.\ref{RS4} (top and center) towards completion. For later times there is some particle assembly because there is a substantial concentration of free pentamers at $t=20,000$. Because the width of the assembly barrier of MLD=9, $N_p=8$ molecules is significantly smaller than that of MLD=19, $N_p=2$ molecule, see again Fig.\ref{RS4}, this leads nearly exclusively to formation of MLD=9, $N_p=8$ particles. This leads to some kinetic selectivity in favor of the MLD=9, $N_p=8$ molecules but the selectivity produced by the two-stage assembly process is quite weak compared to that of the one-stage assembly process. On the other hand, the two-stage assembly scenario very much speeds up the formation of assembled particles: if selectivity is not a central aim then two-stage assembly appears to be more efficient than one-stage assembly.

\pagebreak

\section{Conclusion}
In summary, we have analyzed the kinetic properties of the spanning-tree model for the assembly of viral particles and their selection of RNA molecules. The dependence on the time-dependent occupation probabilities of partial and complete assemblies on the topology and geometry of the spanning-tree molecules that represent the outer part of condensed viral RNA was found to be largely, though not completely, determined by two criteria: the maximum ladder distance (MLD), which is a topological measure of the degree of branching of the spanning tree, and the Wrapping Number ($N_P$), which is a geometrical measure that counts the number of sites that maximally accommodate capsomers. The assembly kinetics is characterized by two time scales: the delay time $t_d$ for the onset of production of particles and the relaxation time $t_r$ for full thermal equilibration. Because of the strong dependence of the relaxation times on the MLD and $N_P$ numbers of the molecules, the assembly kinetics can select for RNA molecules with large $N_P$ and small $MLD$. This is a \textit{purely kinetic effect} that disappears on time-scales large compared to the relaxation time. We carried out numerical assembly competition experiments that showed that this kinetic selectivity can persist over very long time scales.  and, surprisingly, is amplified (i) by supersaturation and (ii) by lower capsid protein to RNA concentration ratios.

The model itself is not a realistic description for any particular virus so a quantitative validation is not possible. However, there are a number of general predictions that are expected to carry over to more realistic models and that can be tested experimentally.  The first is that RNA selectivity should work much better under conditions of moderate supersaturation than under assembly equilibrium conditions. Next, because selective nucleation only ``works" when the assembly activation energy barrier is large compared to the thermal energy, weakening the affinity between capsid proteins by adjusting the pH level should reduce the selectivity. In general, effects that enhance the role of entropy tend to erase selectivity. The third prediction is that RNA selectivity is significantly weakened under a collective assembly scenario, such as the en-masse scenario, though such scenarios do provide a higher yield than protein-by-protein assembly. Finally, according to the model, there should be a strong correlation between RNA selectivity and the degree of internal order of the packaged RNA.  The MS2 virus, whose interior has an ordered internal RNA structure -- under the action of packaging signals -- is expected to have a much higher selectivity than the CCMV virus, whose internal RNA structure is disordered. Consistent with this prediction, the assembly of CCMV is believed to follow an en-masse scenario \cite{Tresset2020}.

There are other cases in cell and molecular biology where kinetic selection is more effective than selection based on thermodynamic equilibrium. A well-known case is the fidelity of DNA duplication during cell division, which is much higher than expected based on thermodynamic equilibrium considerations. Kinetic selection is often associated with the Hopfield proofreading mechanism \cite{hopfield}. In that case, the assembly steps are constantly ``challenged". For the case of DNA duplication this takes place by nuclease activity attempting to break the bond between base-pairs. Since mis-pairing is associated with a weaker bond, the fraction of Watson-Crick paired bonds that survive the challenge is much larger than that of mis-paired bonds. Such a proofreading mechanism intrinsically consumes free energy. Could this apply to the present case? Consider initial formation of small pentamer clusters according to the Boltzmann distribution. Treat an MLD=9, $N_p=8$ assembly as a form of proper pairing and a MLD=19, $N_p=2$ assembly as a form of mis-pairing.  Under conditions of supersaturation, MLD=9, $N_p=8$ assemblies can easily slide down to near-irrevocable completion but clusters on the wide activation barrier of the MLD=19, $N_p=2$ assemblies are frequently challenged against disassembly by thermal fluctuations. While this is not exactly the same as Hopfield proofreading, it is quite similar. The fact that we need the quasi-irreversibility and free energy consumption provided by supersaturation for the mechanism to be efficient strengthens the similarity. 

An aspect of the model that should be improved is the fact that it does not account for conformational fluctuations of the genome molecules, prior to assembly. For the same reason, the model also does not include the ``antenna" effect for the diffusive influx of capsid proteins to partial assemblies \cite{Hu2006}. Another aspect of the model that should be improved in the context of asymmetric reconstructions is a more realistic treatment of the the condensed RNA molecule. Such a generalization would start from a determination of the key interaction sites between capsid proteins and the surface of the enclosed RNA. These sites would span a polyhedron with many more sites than the vertices of a dodecahedron. The spanning tree should reproduce the outer density of the RNA molecule, such as Fig.1. Classifying the set of all spanning trees would be significantly more challenging and require more extensive numerical work.

\begin{acknowledgements}
We would like to thank Alexander Grosberg for drawing our attention to spanning trees and Ioulia Rouzina for introducing us to the concept of selective nucleation. We would like to thank Reidun Twarock, Charles Knobler and William Gelbart for reading a first draft and commenting on it. We also benefitted from discussions with Chen Lin, Zach Gvildys and William Vong. RB would like to thank the NSF-DMR for continued support under CMMT Grant No.1836404.
\end{acknowledgements}

\clearpage

\begin{appendix}
\section{Demonstration that the smallest MLD for spanning trees on the dodecahedron is nine} \label{app:A}
\begin{figure}[htbp]
\begin{center}
\includegraphics[width=2.5in]{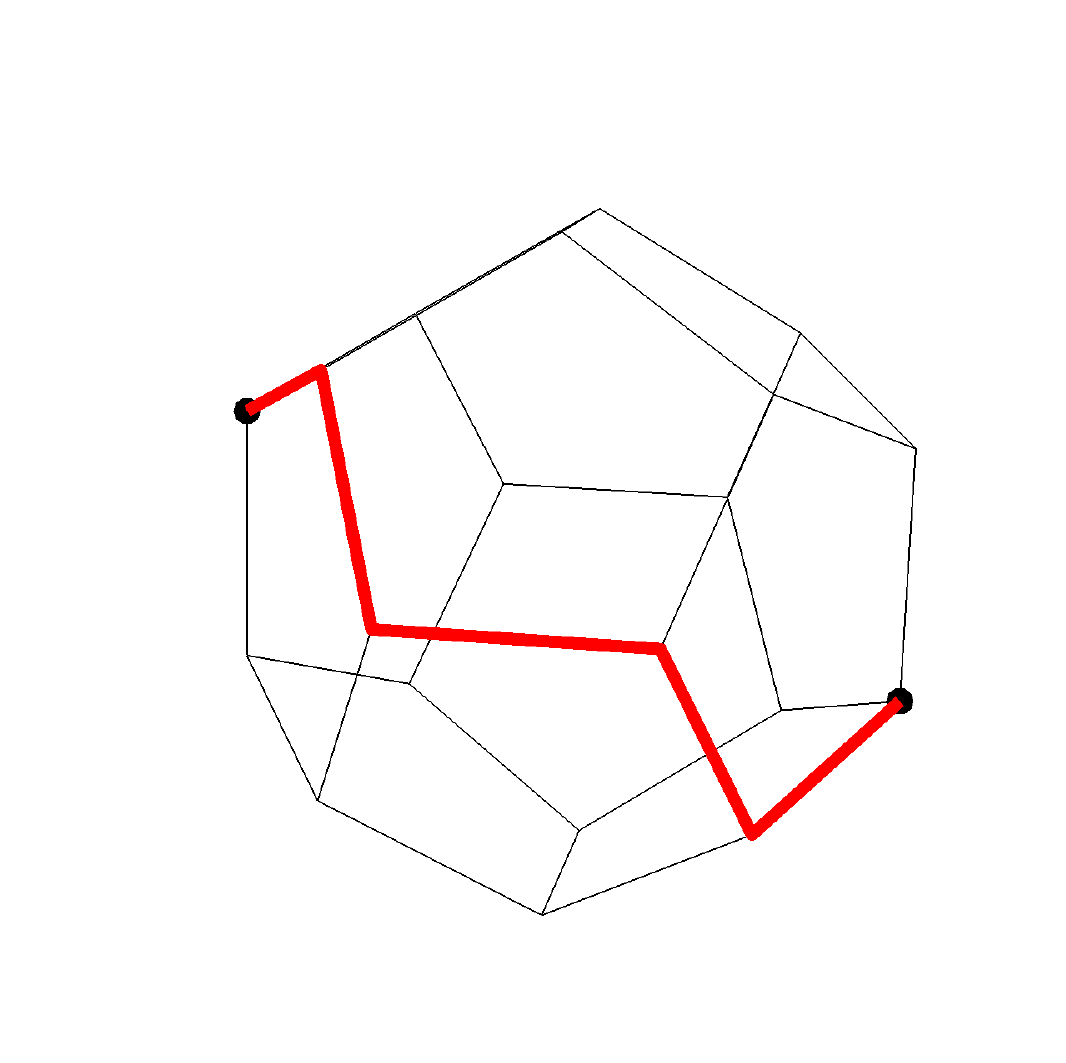}
\caption{Two maximally separated vertices on the dodecahedron and one of the 12 shortest paths consisting of five edges that join them.}
\label{fig:app1}
\end{center}
\end{figure}

We begin by noting that for every vertex on the dodecahedron there is a vertex on the opposite side of the polyhedron that is a ladder distance five away. That is, getting from one of the two vertices to the other requires traversing at least five edges. Figure \ref{fig:app1} shows such a path.
For each such pairs of vertices there are 12 minimal paths. 

\begin{figure}[htbp]
\begin{center}
\includegraphics[width=2.5in]{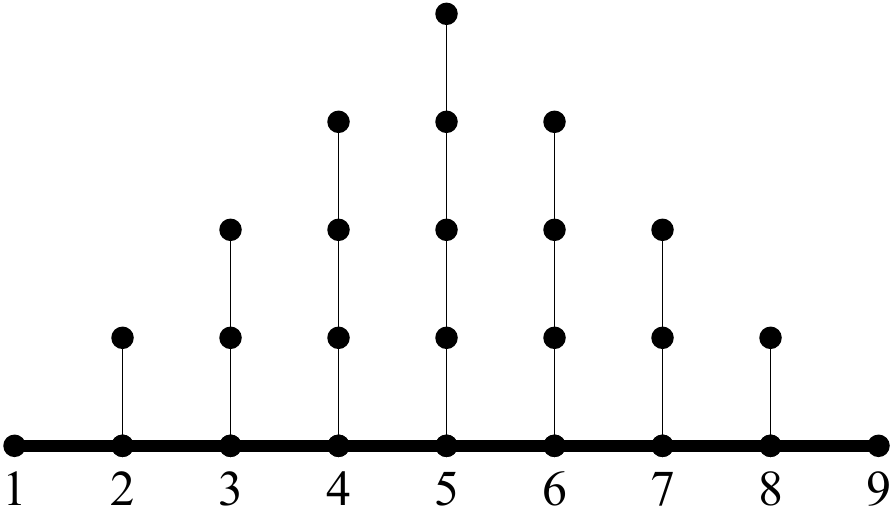}
\caption{A ladder distance 8 path in the hypothetical MLD 8 spanning tree on the dodecahedron. The path is shown as a thick line, and the nine vertices are labeled for easy reference. The thinner vertical lines represent longest allowed paths branching off the ladder distance 8 path. }
\label{fig:app2}
\end{center}
\end{figure}

Now, assume that there is a spanning tree with MLD 8. In such a case, we can pick out a path of ladder distance eight in that tree. All other elements of the tree will consist of trees that branch out from that path. Figure \ref{fig:app2} is a figurative depiction of the path along with the longest allowed branch sprouting off each vertex on that path. The likelihood of branching off those ``side branch'' paths is ignored; such branching does not alter the argument below. 

Consider  first the central vertex on the ladder distance eight path, labeled 5 in Fig.\ref{fig:app2}. The side path with ladder distance four is the longest that can attach to it. A longer path increases the MLD of the tree. Clearly, there is no possibility of reaching a point a ladder distance five from vertex 5 along any path with ladder distance four, so the path shown cannot connect the central vertex to the vertex a distance five away from it.  Next, consider the two sites flanking the central vertices, labeled 4 and 6. Attached to each is the longest possible path branching out from them, Such a path has ladder distance three. If either of these paths reached to the vertex a ladder distance five away from the central vertex, then there would be a ladder distance four (or less) path from that vertex through one of the flanking vertices to the maximally separated vertex, and we know that no such path exists. We can continue this argument to encompass all allowed paths sprouting from vertices on the chosen path. Thus, there is a vertex on the dodecahedron that cannot be a part of the MLD  8 tree containing this path. Consequently no tree with MLD 8 can be a spanning tree on the dodecahedron. The argument above can clearly be applied to the possibility of a spanning tree with MLD less than eight. That there is a spanning tree with MLD 9 is readily established by construction.

\newpage

\section{Boltzmann Distribution}\label{app:B}
In this Appendix we discuss the equilibrium phase behavior of the model assuming the low-temperature Boltzmann Distribution:
\begin{equation}
P_n=\frac{\exp-\Delta F(n)}{Z}
\label{BD}
\end{equation}
for the occupation probabilities. Here $\Delta F(n)= \beta\Delta E(n) - \ln m(n)-n\ln c_f$ is the dimensionless free energy and $Z=\sum_{n=0}^{12}\exp-\Delta F(n)$ the partition sum. We will assume a solution containing only one class of RNA molecules with total concentration $r_t$ as well as pentamers with a total concentration $c_0$. The concentration $r_{n}$ of particles containing $n$ pentamers is then $r_t P_{n}$. Finally, $c_f$ is the concentration of free pentamers and $r_f$ the concentration of unoccupied RNA molecules. Because $m(12)=1$
\begin{equation}
\begin{split}
&r_{12}/r_t=\frac{\exp-\Delta E(12)}{Z}\\&
r_f /r_t= \frac{1}{Z}
\label{c12}
\end{split}
\end{equation} 
for $c_f=1$ (i.e., the reference concentration). Using these relations, it can be checked that
\begin{equation}
\frac{c_f^{12} r_f}{r_{12}}=K
\label{LMA}
\end{equation}
with $K=\exp \Delta E(12)$. This relation has the form of the \textit{Law of Mass Action} (LMA) of physical chemistry with $K$ the dissociation constant. 

Conservation of tree molecules requires that $r_f+\sum_{n=1}^{12} r_n= r_t$, which is assured if the probabilities sum to one $\sum_{n=0}^{12} P_{n} = 1$. Next, conservation of pentamer molecules requires that 
 \begin{equation}
c_f/c_0 = 1 - (D/12)\sum_{n=1}^{12} n P_{n} 
\label{eq:gammadef}
\end{equation}
with $D = 12 r_t/c_0$ the mixing ratio. Recall that if $D=1$ then there are exactly enough pentamers to encapsidate all tree molecules. Because the concentration of free pentamers depends on the Boltzmann distributions of all aggregate sizes, the occupation probabilities for different values of $n$ are coupled. This means that the concentration of assembled particles cannot be obtained from the LMA by itself, but would need to be complemented by similar relations for the concentrations of assembly intermediates. If however the intermediate occupation probabilities can be neglected with respect to $P_0$ and $P_{12}$ then the conservation law for RNA molecules reduces to $P_{0} \simeq (1-P_{12})$ and that of pentamers to $c_f \simeq c_0(1-DP_{12})$. Inserting these two relations into the LMA equation Eq.\ref{LMA} produces an closed-form expression for the concentration $c_f$ of free pentamers and hence of assembled particles:
\begin{equation}
\left(\frac{c_f}{c_0}\right)^{12}\left(\frac{D-1+\frac{c_f}{c_0}}{1-\frac{c_f}{c_0}}\right)=\left(\frac{K}{c_0^{12}}\right)
\label{C1}
\end{equation}
Because $K$ depends only on the assembly energy of complete particles, Eq.\ref{C1} is independent of the class of spanning tree molecules.

\begin{figure}[htbp]
\begin{center}
\includegraphics[width=3in]{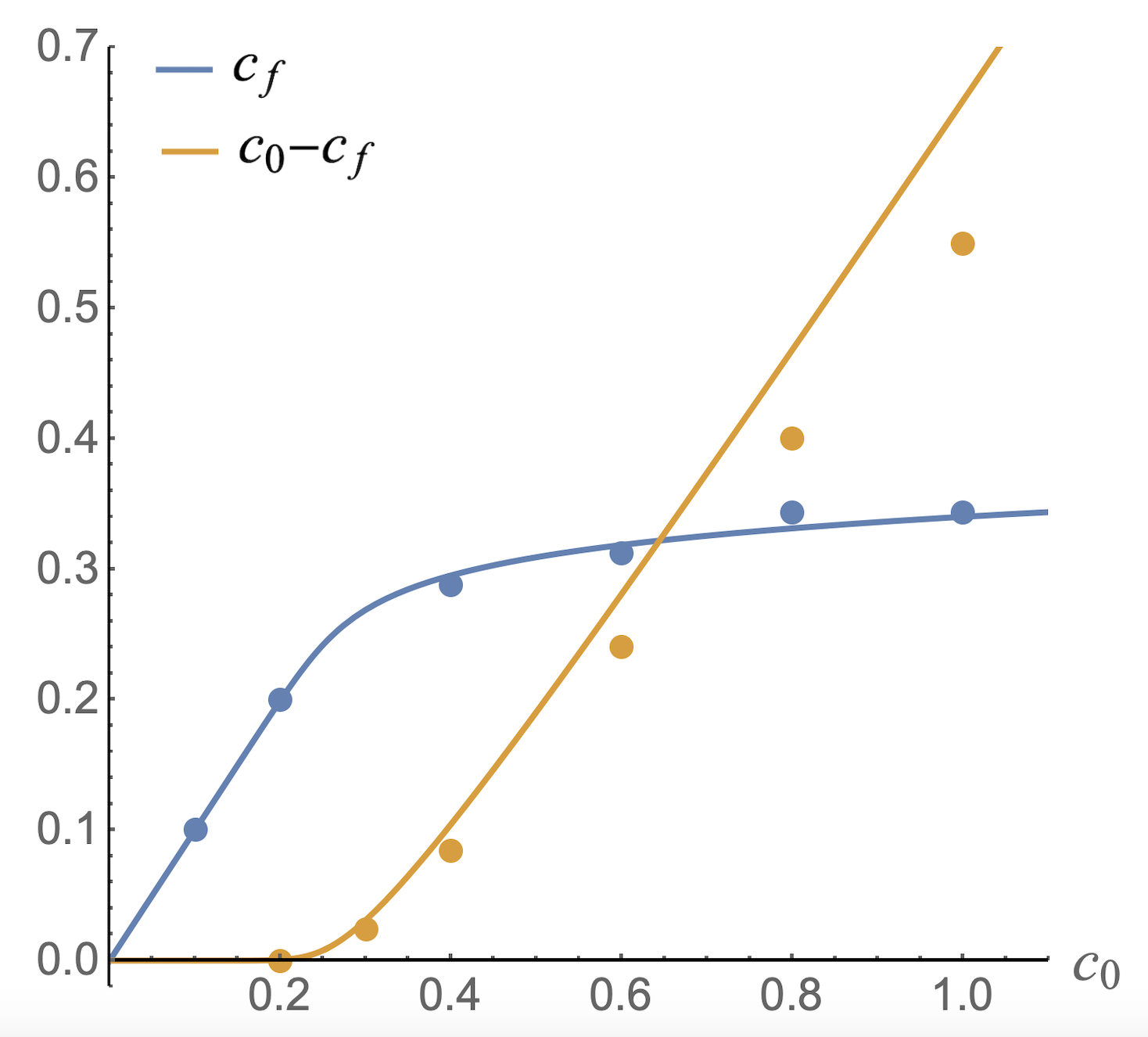}
\caption{Equilibrium self-assembly diagram for class (1) molecules with $\epsilon=-0.2$, $\mu_0=-2.5$, and $D=1$. Horizontal axis: total pentamer concentration $c_0$. Vertical axis: either the free pentamer concentration $c_f$ (blue) or the concentration $c_0-c_f$ of pentamers that are associated with a tree molecule (ochre). Solid lines: solution of Eq.\ref{C1}}
\label{fig:EA}
\end{center}
\end{figure}

A standard diagnostic for self-assembly processes are plots of the concentration of free monomeric building blocks and of assembled particles as a function of the total concentration of building blocks \cite{safran}. Such a plot is shown in Fig.\ref{fig:EA}. The dots show the concentrations of free pentamers in solution and of pentamers that are part of an assembled particle as a function of the total pentamer concentration $c_0$ computed from Eq.\ref{C1}.
For low pentamer concentrations, nearly all pentamers are free in solution and the concentration of free pentamers is close to the total concentration. $c_0$. As $c_0$ increases, the concentration of free pentamers stops increasing and then saturates. Now, the concentration of pentamers that are part of an assembled particle starts to increase, proportional to $c_0$. The transition point between these two regimes is around $c_0=0.2$. This point is known in the soft-matter physics literature as the \textit{critical aggregation concentration} (or CAC) \cite{safran}. The solution of Eq.\ref{C1} that neglects assembly intermediates (solid lines) provides a good approximation.

\begin{figure}[htbp]
\begin{center}
\includegraphics[width=3.5in]{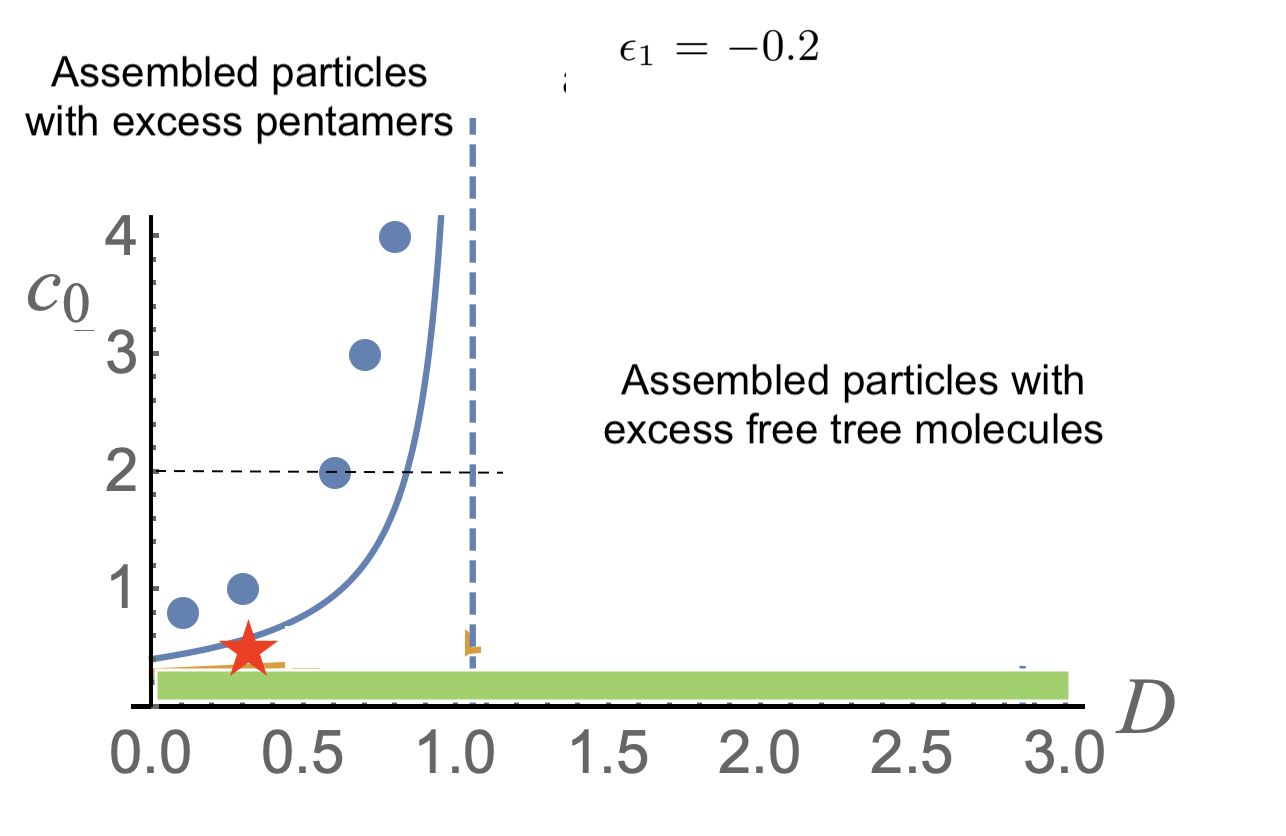}
\caption{Quasi phase-diagram for $\epsilon=-0.2$ and $\mu_0=-2$. Horizontal axis: Depletion factor $D$. Vertical axis: Pentamer concentration $c_0$. Blue dots: points where 95 percent of the genome molecules have been packaged according to the Boltzmann distribution. Solid blue line: computed from Eq.\ref{contour1}. In the green sector there is practically no capsid assembly. The red star marks a possible operating point for viral assembly inside infected cells, just above the CAC, under conditions of excess pentamers in solution.}
\label{fig:PD}
\end{center}
\end{figure}

A second way to display self-assembly measurements under equilibrium conditions is in the form of a quasi phase-diagram that shows the dominant type of assembly as a function of thermodynamic parameters \footnote{Since a virus is a system of limited size, true phase transitions are not possible.}. For viral assembly, the protein and RNA concentrations are a natural choice for such a phase-diagram. For the case of the spanning-tree model, we will use the pentamer concentration $c_0$ and the mixing ratio $D$ as thermodynamic parameters. The blue dots in Fig.\ref{fig:PD} show points in a $c_0$ vs $D$ diagram where 95 percent of the spanning trees are fully encapsidated. 
To the right of the blue dots, most tree molecules are encapsidated and coexist with excess free pentamers. To the left of the blue dots most pentamers are part of assembled particles and coexist with excess free tree molecules. The blue dots can be viewed as ``optimal mixing states" that minimize excess free pentamers and excess tree molecules. For high pentamer concentrations, the line of blue dots approaches $D=1$, the stoichiometric ratio. 

If one neglects assembly intermediates then it can be shown from Eq.\ref{C1} that the relation $c_0(D)$ for 95 percent occupancy is a hyperbola in the $c_0-D$ plane:
\begin{equation}
c_0(D)\simeq\frac{1}{(1-D\thinspace P_{12})}\left(\frac{K P_{12}}{1-P_{12}}\right)^{1/12}
\label{contour1}
\end{equation}
with $P_{12}=0.95$. The hyperbola diverges at $D=1/P_{12}$, which is close to one for a 95 percent packaging fraction. It shifts to smaller values of $D$ as the pentamer concentration $c_0$ is reduced with $c_0(D)$ always larger than $K^{1/12}$. The green sector of Fig.\ref{fig:PD} is below the CAC. 

For $\left(\frac{c_0^{12}}{K}\right)$ small compared to one, the equation has a solution with $c_f$ close to $c_0$ given by:
\begin{equation}
\frac{c_f}{c_0}\simeq 1 - \frac{D c_0^{12}}{K}
\end{equation}
For $\left(\frac{c_0^{12}}{K}\right)$ large compared to one and mixing ratio $D$ larger than one, the equation has a solution with $c_f$ independent of $c_0$:
\begin{equation}
{c_f}\simeq \left(\frac{K}{D-1}\right)^{1/12}
\end{equation}
Finally, for $\left(\frac{c_0^{12}}{K}\right)$ large compared to one but the mixing ratio $D$ less than one, the equation has a different solution with $c_f$ independent of $c_0$:
\begin{equation}
\frac{c_f}{c_0}\simeq 1 - D +\frac{K}{c_0^{12}}\frac{D}{(1-D)^{1/2}}
\end{equation}
There is thus a change in regimes near the point where the mixing ratio is equal to one. For the special case that $D=1$, the LMA equation reduces to
\begin{equation}
\left(\frac{c_f}{c_0}\right)^{13}\left(\frac{1}{1-\frac{c_f}{c_0}}\right)\simeq\left(\frac{K}{c_0^{12}}\right)
\label{B}
\end{equation}
\end{appendix}

\end{document}